\documentclass[showpacs,nofootinbib,showkeys,eqsecnum,prd,aps,preprint]{revtex4}

\usepackage[english]{babel}
\usepackage[cp1251]{inputenc}
\usepackage[T1]{fontenc}

\usepackage{amssymb}
\usepackage{amsmath}
\usepackage{amsfonts}
\usepackage{latexsym}
\usepackage{mathrsfs}
\usepackage{bm}
\usepackage{tensor}
\usepackage{hyperref}

\begin{document}

\title{Vacuum instability in QED with an asymmetric $x$-step. New example of
	exactly solvable case}

\author{A. I. Breev }
\email{breev@mail.tsu.ru}
\affiliation{Department of Theoretical Physics, Tomsk State University Novosobornaya
	Sq. 1, 634050 Tomsk, Russia}

\author{S. P. Gavrilov }
\email{gavrilovsergeyp@yahoo.com, gavrilovsp@herzen.spb.ru}
\affiliation{Herzen State Pedagogical University of Russia, Moyka embankment 48, 191186, St. Petersburg, Russia}

\author{D. M. Gitman }
\email{dmitrygitman@hotmail.com}
\affiliation{P.N. Lebedev Physical Institute, 53 Leninskiy ave., 119991 Moscow, Russia;}
\affiliation{Institute of Physics, University of S\~{a}o Paulo, Rua do Mat\~{a}o, 1371, CEP 05508-090, S\~{a}o Paulo, SP, Brazil.}

\begin{abstract}
We present a new exactly solvable case in strong-field $QED$ with one-dimensional step potential ($x$-step). The corresponding $x$-step is
given by an analytic asymmetric with respect to the axis $x$ reflection
function. The step can be considered as a certain analytic "deformation" of
the symmetric Sauter field. Moreover, it can be treated as a new
regularization of the Klein step field. We study the vacuum instability
caused by this $x$-step in the framework of a nonperturbative approach to
strong-field $QED$. Exact solutions of the Dirac equation used in the
corresponding nonperturbative calculations, are represented in the form of
stationary plane waves with special left and right asymptotics and
identified as components of initial and final wave packets of particles. We
show that in spite of the fact that the symmetry with respect to positive
and negative bands of energies is broken, distribution of created pairs and
other physical quantities can be expressed via elementary functions. We
consider the processes of transmission and reflection in the ranges of the
stable vacuum and study physical quantities specifying the vacuum
instability. We find the differential mean numbers of electron-positron
pairs created from the vacuum, the components of current density and
energy-momentum tensor of the created electrons and positrons leaving the
area of the strong field under consideration. Besides, we study the
particular case of the particle creation due to a weakly inhomogeneous
electric field and obtain explicitly the total number, the current density
and energy-momentum tensor of created particles. Unlike the symmetric case
of the Sauter field the asymmetric form of the field under consideration
causes the energy density and longitudinal pressure of created electrons to
be not equal to the energy density and longitudinal pressure of created
positrons.
\end{abstract}

\pacs{03.65.Pm, 12.20.-m, 23.20.Ra}

\keywords{Pair creation, Schwinger effect, Dirac equation, exact solutions \\ Mathematics Subject Classification 2010: 81Q05, 81V10, 33C50}

\maketitle

\section{Introduction\label{S1}}

The Schwinger effect, that is, creation of charged particles from the vacuum
by strong external electric-like and gravitational fields (the vacuum
instability) has been attracting attention already for a long time; see,
e.g., monographs ~\cite{BirDav82,GMR85,GriMaM,FGS}. This is a
nonperturbative effect of quantum field theory ($QFT$), which has not yet
received a convincing experimental confirmation. However, recent progress in
laser physics allows one to hope that the vacuum instability will be
experimentally observed in the near future even in laboratory conditions.
Recently, this, as well as the real possibility of observing an analogue of
the Schwinger effect in condensed matter physics (in the graphene,
topological insulators, 3D Dirac and Weyl semimetals, antiferromagnets,
etc.) has increased theoretical interest in the problem and led to the
development of various analytical and numeric approaches, see recent reviews 
\cite{VafVis14,GelTan16,Adv-QED22,Progr23}. From general quantum theory
point of view, the most clear formulation of the problem of particle
production from the vacuum by external fields is formulated for
time-dependent external electric fields that are switched on and off at
infinitely remote times $t\rightarrow \pm \infty$, respectively. The
idealized problem statement described above was considered for uniform
time-dependent external electric fields. Such kind of external fields are
called the $t$-electric potential steps ($t$-steps). A complete
nonperturbative with respect to the external background formulation of
strong-field $QED$ with such external fields was developed in Refs. \cite%
{FGS,Gitma77}; it is based on the existence of exact solutions of the Dirac
equation with time dependent external field (more exactly, complete sets of
exact solutions). However, there exist many physically interesting
situations in high-energy physics, astrophysics, and condensed matter where
external backgrounds formally are time-independent. In our works; see Refs. 
\cite{GavGi16,GavGi20,BGavGi23}, a nonperturbative approach in $QED$ with
the so-called $x$-potential steps, or simply $x$-steps, was developed. The $x
$-steps represent time-independent inhomogeneous electric-like external
fields of a constant direction. The latter approach is based on the
existence of special exact solutions of the Dirac or Klein-Gordon equations
with corresponding $x$-steps. In cases when such solutions can be found and
all the calculations can be done analytically, we refer to these cases as to
exactly solvable ones. Sauter potential and the Klein step, considered in
the pioneer works \cite{Klein27,Sauter31a,Sauter-pot}, belong to the class
of exactly solvable cases. Initially they were considered in the framework
of the relativistic quantum mechanics, which gave rise to a rather
long-lasting discussion about the Klein paradox (a detailed historical
review can be found in Refs. \cite{DomCal99,HansRavn81}). In the work \cite%
{GavGi16} it was pointed out that this paradox and other misunderstandings
in considering quantum effects in fields of strong $x$-steps can be
consistently solved as many particle effects of the $QFT$ ($QED$) with
unstable vacuum. Recently, a number of new exactly solvable cases were
presented and studied in detail in the framework of general approach \cite%
{GavGi16,GavGi20}. Particularly, interesting are the cases of a constant
electric field between two capacitor plates ($L$-constant electric field) 
\cite{GavGit16b}, a field of a piecewise form of continuous exponential
functions \cite{GavGitSh17}, and a piecewise and a continuous configuration
of an inverse-square step \cite{AdoGavGit20}. Exactly solvable cases are
interesting not only in themselves, but also due to the fact that they allow
you to develop and test new approximate and numerical methods for
calculating quantum effects in strong-field $QFT$. One can find a
number of application of these exactly solvable cases in high-energy physics
and condensed matter physics; see, e.g., \cite%
{GavGit13,AbrZub16,AdoHeGG21,AdoGavGit23}.

In this article, we present a new exactly solvable case for strong-field $%
QED $ with $x$-step. For the generality, the field is considered in $d=D+1$
- dimensional Minkowski space-time, parametrized by the coordinates $%
X=\left( t,\mathbf{r}\right) $, $\mathbf{r}=\left( x^{1}=x,\mathbf{r}%
_{\perp }\right) $, $\mathbf{r}_{\perp }=x^{2},\ldots ,x^{D}$. The electric
field is constant and has only one component along the $x$-axis, $\mathbf{E}%
\left( X\right) =\left( E^{1}\left( x\right) =E\left( x\right)
,0,...,0\right) $. The field is given by a step potential $A_{0}(x)$, so
that $E\left( x\right) =-A_{0}^{\prime }\left( x\right) $.

We note that among the above exactly solvable cases only the Sauter electric
field is given by an analytic function,%
\begin{eqnarray}
&&A_{0}^{(\mathrm{Sauter})}(x;L,E_{S})=-LE_{S}\tanh \left( x/L\right) , 
\notag \\
&&E_{(\mathrm{Sauter})}(x;L,E_{S})=E_{S}\cosh ^{-2}\left( x/L\right)
,\;E_{S}>0,\;L>0.  \label{1.4}
\end{eqnarray}%
This field reaches its maximum value at $x=0$ and is symmetric with respect
to the origin. Unlike the above mentioned cases given by piecewise smooth
x-steps, physical quantities calculated for the analytic Sauter field are
presented by elementary functions, which makes this case especially
convenient for physical interpretations. Here we present a new example of
exactly solvable case in which the external field is given by the following
analytic function:%
\begin{eqnarray}
&&A_{0}(x)=\frac{\sigma E_{0}}{\sqrt{1+\exp \left( \frac{x}{\sigma }\right) }%
},\quad E_{0}>0,\ \sigma >0,\   \notag \\
&&E(x)=\frac{E_{0}}{8}\sqrt{1+\exp \left( \frac{x}{\sigma }\right) }\cosh
^{-2}\left( \frac{x}{2\sigma }\right) \ .  \label{1.2}
\end{eqnarray}%
The potential energy of an electron (with the charge $q=-e,$ $e>0$) is $%
U\left( x\right) =-eA_{0}\left( x\right) $. It tends to different in the
general case constants values $U\left( -\infty \right) $ and $U\left(
+\infty \right) $ as $x\rightarrow -\infty $ and $x\rightarrow +\infty $,
respectively,%
\begin{equation}
U\left( -\infty \right) \equiv U_{\mathrm{L}}=-eE_{0}\sigma ,\ \ U\left(
+\infty \right) \equiv U_{\mathrm{R}}=0\,.  \label{U}
\end{equation}%
The magnitude $\delta U$ of the potential step is given by the difference%
\begin{equation}
\delta U=U_{\mathrm{R}}-U_{\mathrm{L}}=eE_{0}\sigma  \label{dU}
\end{equation}%
Note $\delta U$ is equal to the increment of kinetic energy, if the particle
retains the direction of motion and moves in the direction of acceleration,
and if toward the opposite, then this increment changes sign. Depending on
the magnitude $\delta U$, the step is called noncritical or critical one,
see \cite{GavGi16},%
\begin{equation}
\begin{array}{l}
\delta U<\delta U_{c}=2m\,,\text{ \textrm{noncritical step}} \\ 
\delta U>\delta U_{c},\text{ \ \ \ \ \ \ \ \ \ \textrm{critical step}}%
\end{array}%
\,.  \label{cr}
\end{equation}%
If the magnitude $\delta U$ is large enough, the particle production from
the vacuum could be essential.

The electric field (\ref{1.2}) differs from the Sauter field (\ref{1.4}) at $%
L=2\sigma $ by the presence of an additional term in Eq. (\ref{1.2}),%
\begin{equation}
E(x)=\sqrt{1+\exp \left( \frac{x}{\sigma }\right) }E_{(\mathrm{Sauter}%
)}\left( x;2\sigma ,\frac{E_{0}}{8}\right) .  \label{1.5}
\end{equation}%
There is no symmetry of the field $E(x)$ with respect to the point $x_{%
\mathrm{M}}=\sigma \ln 2$, in which the field has the maximum value $E_{\max
}=E_{0}/(3\sqrt{3})$. While the Sauter field exhibits this symmetry with
respect to the point of its maximum. Moreover, they generally increase in a
similar way, but the Sauter field decreases faster. One can say that field (%
\ref{1.2}) for a given value of the parameter $\sigma $ is a certain
"deformation" of the Sauter field, which turns on at $x>0$ and turns off at $%
x\rightarrow +\infty $. Finally, we note that distributions of created pairs
by the Sauter field are symmetric with respect of the energy $p_{0}$. The
latter symmetry is not inherent in realistic asymmetric fields. We
demonstrate that in the case of the asymmetric analytic field (\ref{1.2}),
with broken symmetry with respect to $p_{0}$, the distributions of created
pairs and other physical quantities can be still expressed in terms of
elementary functions.

The article is organized as follows: In Sect. \ref{S2}, we construct exact
solutions of the Dirac equation with a new example of $x$-step given by a
analytic asymmetric function. These solutions are presented in the form of
stationary plane waves with special left\ and right\ asymptotics and
identified as components of initial and final wave packets of particles and
antiparticles. We find coefficients of mutual decompositions of the initial
and final solutions. In Sec. \ref{S3}, we consider the processes of
transmission and reflection in ranges of the stable vacuum. In Sec. \ref{S4}%
, we calculate physical quantities specifying the vacuum instability. We
find differential mean numbers of electron-positron pairs created from the
vacuum, as well as components of current density and energy-momentum tensor
of the created electrons and positrons leaving the area of the strong
external field. In Sec. \ref{S5}, we consider a particular case of the
particle creation due to a weakly inhomogeneous electric field and obtain
explicitly the total number, current density and energy-momentum tensor of
the\ created particles. A new regularization of the Klein step is considered
in Sec. \ref{S6}, which is used then in calculating the corresponding vacuum
instability. Section \ref{S7} contains some concluding remarks. In Appendix %
\ref{Ap}, we describe briefly basic elements of a nonperturbative approach
to $QED$ with $x$-steps. In Appendix \ref{Ap2}, we show that the density of
created pairs and the probability of the vacuum to remain a vacuum obtained
from exact formulae for the slowly varying field in the leading-term
approximation are in agreement with results following in the framework of a
locally constant field approximation (LCFA). In Appendix \ref{A1}, we list
some useful properties of hypergeometric functions. We use the system of
units, where $c=\hbar =1$.

\section{Solutions of Dirac equation with asymmetric potential $x$-step\label%
{S2}}

\subsection{General solution}

Let us consider the Dirac equation with a $x$-step in the Hamiltonian form:%
\begin{eqnarray}
&&i\partial _{0}\psi \left( X\right) =\hat{H}\psi \left( X\right) \,,  \notag
\\
&&\hat{H}=\gamma ^{0}\left( -i\gamma ^{j}\partial _{j}+m\right) +U\left(
x\right) \,,\ \ j=1,\ldots ,D\,.  \label{2.6}
\end{eqnarray}%
The Dirac spinor $\psi \left( X\right) $ has $2^{\left[ d/2\right] }$
components, $\left[ d/2\right] $ denotes the integer part of $d/2$, and $%
\gamma ^{\mu }$ are $2^{\left[ d/2\right] }\times 2^{\left[ d/2\right] }$
Dirac matrices in $d$ dimensions, $\left[ \gamma ^{\mu },\gamma ^{\nu }%
\right] _{+}=2\eta ^{\mu \nu }\,\,$, and $U\left( x\right) =-eA_{0}\left(
x\right) ,$ where $A_{0}(x)$ is given by Eq. (\ref{1.2}).

There exist solutions of Eq. (\ref{2.6}) in the form of stationary plane
waves propagating along the space-time directions $t$ and $\mathbf{r}_{\perp
}$. In this case the Dirac spinors labeled by quantum numbers $n$ have the
form:%
\begin{eqnarray}
&&\psi _{n}\left( X\right) =\exp \left( -ip_{0}t+i\mathbf{p}_{\perp }\mathbf{%
r}_{\perp }\right) \psi _{n}\left( x\right) \,,\ \ n=\left( p_{0},\mathbf{p}%
_{\perp },\sigma \right) \,,  \notag \\
&&\psi _{n}\left( x\right) =\left\{ \gamma ^{0}\left[ p_{0}-U\left( x\right) %
\right] +i\gamma ^{1}\partial _{x}-\boldsymbol{\gamma }_{\perp }\mathbf{p}%
_{\perp }+m\right\} \varphi _{n}\left( x\right) v_{\chi ,\sigma }\,,
\label{2.8}
\end{eqnarray}%
where the spinors $\psi _{n}\left( x\right) $ and the scalar functions $%
\varphi _{n}\left( x\right) $ depend exclusively on $x$ while $v_{\chi
,\sigma }$ is a set of constant orthonormalized spinors, satisfying the
following conditions:%
\begin{eqnarray}
\gamma ^{0}\gamma ^{1}v_{\chi ,\sigma } &=&\chi v_{\chi ,\sigma },\ \
v_{\chi ,\sigma }^{\dag }v_{\chi ^{\prime },\sigma ^{\prime }}=\delta _{\chi
,\chi ^{\prime }}\delta _{\sigma ,\sigma ^{\prime }}\ ,  \notag \\
\chi &=&\pm 1,\ \ \sigma =\left( \sigma _{s}=\pm 1\,,\ s=1,2,...,\left[ d/2%
\right] -1\right)  \label{2.7}
\end{eqnarray}%
Quantum numbers $s$\ and $\chi $\ describe the spin polarization (if $d\leq
3 $\ there are no spin degrees of freedom that are described by the quantum
numbers{\large \ }$s${\large ). }Solutions $\psi _{n}^{\left( \chi \right)
}\left( X\right) $ and $\psi _{n}^{\left( \chi ^{\prime }\right) }\left(
X\right) $ given by Eqs. (\ref{2.8}) that differ only by values of $\chi $
are linearly dependent\emph{\ }if\emph{\ }$d>3$\emph{.} Therefore, it
suffices to work with solutions corresponding to one of possible values of $%
\chi $, and sometimes we omit the subscript $\chi $, supposing that the spin
quantum number $\chi $ is fixed in a certain way. Due to the same reason,
there exists, in fact, only $J_{(d)}=2^{[d/2]-1}$ different spin states
(labeled by quantum numbers $\sigma $) for a given set of $p_{0},\mathbf{p}%
_{\bot }$. Substituting Eq. (\ref{2.8}) into Eq. (\ref{2.6}), one finds that
scalar functions $\varphi _{n}\left( x\right) $ obey the following
second-order ordinary differential equation:%
\begin{equation}
\left\{ \frac{d^{2}}{dx^{2}}+\left[ p_{0}-U\left( x\right) \right] ^{2}-\pi
_{\perp }^{2}+i\chi \partial _{x}U\left( x\right) \right\} \varphi
_{n}\left( x\right) =0\,,\ \pi _{\perp }=\sqrt{\mathbf{p}_{\perp }^{2}+m^{2}}%
.  \label{2.10}
\end{equation}

Solutions of a similar type of equation%
\begin{equation}
\left\{ \frac{d^{2}}{dt^{2}}+\left[ p_{x}-\tilde{U}\left( t\right) \right]
^{2}+\pi _{\perp }^{2}-i\tilde{\chi}\partial _{t}\tilde{U}\left( t\right)
\right\} \tilde{\varphi}_{n}\left( t\right) =0,  \label{t-eq}
\end{equation}%
where $\tilde{U}\left( t\right) =-A_{x}\left( t\right) $, were recently
found in Ref. \cite{BrGavGitSh21}\textrm{\ }for the case of a
nonperturbative treatment of the vacuum instability due to\emph{\ }a
time-dependent electric field $E\left( t\right) $, given by the potential:%
\begin{equation*}
A_{x}\left( t\right) =\frac{\sigma E_{0}}{\sqrt{1+\exp \left( t/\sigma
\right) }}\ .
\end{equation*}%
It is quite obvious that equation (\ref{2.10}) can be obtained from equation
(\ref{t-eq}) by a substitution%
\begin{equation*}
t\rightarrow x,\;p_{x}\rightarrow p_{0},\;\pi _{\perp }^{2}\rightarrow -\pi
_{\perp }^{2},\;\tilde{\chi}\rightarrow -\chi .
\end{equation*}%
Therefore, solutions of equation (\ref{2.10}) can be obtained from solutions
of equation (\ref{t-eq}) using the same substitution. As the result, the
general solution of Eq. (\ref{2.10}) can be represented as a linear
combination of the functions $\varphi _{n,i}\left( x\right) $,%
\begin{eqnarray}
&&\varphi _{n,i}\left( x\right) =\ \left. \left( 1+z\right) ^{i\alpha
_{1}}\left( 1-z\right) ^{i\alpha _{2}}\hat{M}_{n}w_{n,i}\left( \frac{z+1}{2}%
\right) \right\vert _{z=z(x)}\ ,  \notag \\
&&\hat{M}_{n}=\frac{bz-(\alpha _{1}-\alpha _{2}+2\chi eE_{0}\sigma ^{2})}{%
\left( ia-1\right) b}\frac{d}{dz}+1\ ,\ z(x)=\sqrt{1+\exp \left( \frac{x}{%
\sigma }\right) }\ ,  \notag \\
&&a=\alpha _{1}+\alpha _{2}-\sqrt{2\left( \alpha _{1}^{2}+\alpha
_{2}^{2}\right) -(2eE_{0}\sigma ^{2})^{2}},\ \ b=\alpha _{1}+\alpha _{2}+%
\sqrt{2\left( \alpha _{1}^{2}+\alpha _{2}^{2}\right) -(2eE_{0}\sigma
^{2})^{2}}\ ,  \notag \\
&&\alpha _{1}=\sigma \sqrt{\pi _{\perp }^{2}-(p_{0}-eE_{0}\sigma )^{2}},\
\alpha _{2}=\sigma \sqrt{\pi _{\perp }^{2}-(p_{0}+eE_{0}\sigma )^{2}}\ .
\label{2.15}
\end{eqnarray}%
In the above combination, we use two pairs of linearly independent solutions 
$w_{n,i}\left( \xi \right) $ with additional indices $i=1,\ldots ,4$. The
first pair reads: 
\begin{eqnarray}
&&w_{n,1}\left( \xi \right) =\xi ^{-(ia-1)}F\left( ia-1,i(a-2\alpha
_{1});2i\alpha _{2};1-\xi ^{-1}\right) \ ,  \notag \\
&&w_{n,2}\left( \xi \right) =\xi ^{i(a-2\alpha _{1})-1}(1-\xi )^{1-2i\alpha
_{2}}F\left( i(2\alpha _{1}-a),-ia;2(1-i\alpha _{2});1-\xi ^{-1}\right) \ ,
\label{2.13}
\end{eqnarray}%
where $F\left( \alpha ,\beta ;\gamma ;\xi \right) $ are Gaussian
hypergeometric functions \cite{Bateman}. Solutions (\ref{2.13}) are
well-defined in a vicinity of the singular point $\xi =1,$ which corresponds
to $x\rightarrow -\infty $. The second pair reads:%
\begin{eqnarray}
&&w_{n,3}\left( \xi \right) =\left( -\xi \right) ^{-(ia-1)}F\left(
ia-1,i(a-2\alpha _{1})-1;i(a-b);\xi ^{-1}\right) \ ,  \notag \\
&&w_{n,4}\left( \xi \right) =\left( -\xi \right) ^{-ib}F\left(
ib,i(b-2\alpha _{1})+1;i(b-a)+2;\xi ^{-1}\right) \ .  \label{2.14}
\end{eqnarray}%
Solutions (\ref{2.14}) are well-defined in a vicinity of the singular point $%
\xi =\infty $, which corresponds to $x\rightarrow +\infty $. Using functions
(\ref{2.13}) and (\ref{2.14}) one can construct four complete sets $\varphi
_{n,i}\left( x\right) $, $i=1,2,3,4$, of solutions of equation (\ref{2.10}).

\subsection{Solutions with special left\ and right\ asymptotics}

Unlike the explicitly time-dependent solutions of Eq. (\ref{t-eq}) the
solutions given by Eqs. (\ref{2.8}) and (\ref{2.15}) are stationary plane
waves. In the treatment of the vacuum instability in the $x$-case under
consideration, they describe qualitatively different cases depending on the
ranges of quantum numbers. By this reason their role in calculations and
interpretation of physics of the vacuum instability in the $x$-case is quite
different compared to similar in appearance the above cited time-dependent
solutions. Further, we carry out a nonperturbative study of the vacuum
instability using solutions (\ref{2.15}) within the framework of the
approach formulated in Refs. \cite{GavGi16,GavGi20}.

Due to local properties of equation (\ref{2.10}) at $x\rightarrow \mp \infty 
$ (where the electric field is zero), the scalar functions $\varphi
_{n}\left( x\right) $ have definite left \textquotedblleft \textrm{L}%
\textquotedblright\ and right \textquotedblleft \textrm{R}%
\textquotedblright\ asymptotics:%
\begin{eqnarray}
&&\ _{\zeta }\varphi _{n}\left( x\right) =\ _{\zeta }\mathcal{N}e^{i\zeta
\left\vert p^{\mathrm{L}}\right\vert x}\ \text{\textrm{as}}\ x\rightarrow
-\infty \,,  \notag \\
&&\ ^{\zeta }\varphi _{n}\left( x\right) =\ ^{\zeta }\mathcal{N}e^{i\zeta
\left\vert p^{\mathrm{R}}\right\vert x}\text{ \textrm{as}}\ \ x\rightarrow
+\infty \,.  \label{2.16}
\end{eqnarray}%
Here $_{\zeta }\mathcal{N}$ and$\ ^{\zeta }\mathcal{N}$ are some
normalization constants, and $p^{\mathrm{L/R}}=\zeta \left\vert p^{\mathrm{%
L/R}}\right\vert $, $\zeta =\pm =\mathrm{sgn}\left( p^{\mathrm{L}}\right) =%
\mathrm{sgn}\left( p^{\mathrm{R}}\right) $, denotes real asymptotic momenta
along the $x$-axis,%
\begin{equation}
\left\vert p^{\mathrm{L/R}}\right\vert =\sqrt{\pi _{0}\left( \mathrm{L/R}%
\right) ^{2}-\pi _{\perp }^{2}},\ \ \pi _{0}\left( \mathrm{L/R}\right)
=p_{0}-U_{\mathrm{L/R}}\,,  \label{2.12}
\end{equation}%
where $U_{\mathrm{L/R}}$ is given by Eq. (\ref{U}). Then for the
corresponding Dirac spinors the following relations hold:%
\begin{eqnarray}
\hat{p}_{x}\ _{\zeta }\psi _{n}\left( X\right) &=&\zeta \left\vert p^{%
\mathrm{L}}\right\vert \ _{\zeta }\psi _{n}\left( X\right) \ \text{\textrm{as%
}}\ x\rightarrow -\infty \,,  \notag \\
\hat{p}_{x}\ ^{\zeta }\psi _{n}\left( X\right) &=&\zeta \left\vert p^{%
\mathrm{R}}\right\vert \ ^{\zeta }\psi _{n}\left( X\right) \ \text{\textrm{as%
}}\ x\rightarrow +\infty \,.  \label{2.17}
\end{eqnarray}

Note that the electric field under consideration can be neglected at
sufficiently big $\left\vert x\right\vert $, let say, in the macroscopic
regions $S_{\mathrm{L}}$\ on the left of $x=x_{\mathrm{L}}<0$\ and $S_{%
\mathrm{R}}$\ on the right of\ $x=x_{\mathrm{R}}>0$.\emph{\ }To this end one
can choose finite $x_{\mathrm{L/R}}$\ such that 
\begin{equation}
1-\frac{U\left( x_{\mathrm{R}}\right) -U\left( x_{\mathrm{L}}\right) }{%
\delta U}\ll 1.  \label{asympt}
\end{equation}%
\emph{\ }We assume that the asymptotic behavior (\ref{2.17}) is sufficiently
good approximation for all the functions $\ _{\zeta }\psi _{n}\left(
X\right) $\ and $^{\zeta }\psi _{n}\left( X\right) $\ if $x<x_{\mathrm{L}}$
and $x>x_{\mathrm{R}}$, which means that particles are free in the regions%
\emph{\ }$S_{\mathrm{L}}$ and $S_{\mathrm{R}}$.

Nontrivial sets of Dirac spinors $\left\{ \ _{\zeta }\psi _{n}\left(
X\right) \right\} $ and $\left\{ \ ^{\zeta }\psi _{n}\left( X\right)
\right\} $, that are key elements of the above mentioned approach, do exist
for the quantum numbers $n$ satisfying the conditions:%
\begin{equation}
\pi _{0}\left( \mathrm{L/R}\right) ^{2}>\pi _{\perp }^{2}\Leftrightarrow
\left\{ 
\begin{array}{l}
\pi _{0}\left( \mathrm{L/R}\right) >\pi _{\perp }\, \\ 
\pi _{0}\left( \mathrm{L/R}\right) <-\pi _{\perp }\,%
\end{array}%
\right. \,.  \label{2.19}
\end{equation}%
Note that $\pi _{0}\left( \mathrm{L}\right) >\pi _{0}\left( \mathrm{R}%
\right) $. As a result of these inequalities, the complete set of the
quantum numbers $n$ can be divided in some ranges $\Omega _{k}$, where the
index $k$ labels the ranges and the corresponding quantum numbers, $n_{k}\in
\Omega _{k}$. For critical steps, $\delta U>\delta U_{c}$, there are five
ranges of the quantum numbers, $\Omega _{k}$, $k=1,...,5$, where the
solutions$\ $\emph{\ }$_{\zeta }\psi _{n}\left( X\right) $\emph{\ }and\emph{%
\ }$^{\zeta }\psi _{n}\left( X\right) $\emph{\ }have similar\emph{\ }forms
and properties for given perpendicular momenta $p_{\perp }$\ and any spin
polarizations $\sigma $\emph{.} The ranges $\Omega _{1}$ and $\Omega _{5}$
are characterized by energies bounded from the below, $\Omega _{1}=\left\{
n:p_{0}\geq U_{\mathrm{R}}+\pi _{\perp }\right\} ,$ and by energies bounded
from the above $\Omega _{5}=\left\{ n:p_{0}\leq U_{\mathrm{L}}-\pi _{\perp
}\right\} $. The ranges $\Omega _{2}$ and $\Omega _{4}$ are characterized by
bounded energies, namely \newline
$\Omega _{2}=\left\{ n:U_{\mathrm{R}}-\pi _{\perp }<p_{0}<U_{\mathrm{R}}+\pi
_{\perp }\right\} $ and $\Omega _{4}=\left\{ n:U_{\mathrm{L}}-\pi _{\perp
}<p_{0}<U_{\mathrm{L}}+\pi _{\perp }\right\} $ if $\delta U\geq 2\pi _{\perp
}$ or $\Omega _{2}=\left\{ n:U_{\mathrm{L}}+\pi _{\perp }<p_{0}<U_{\mathrm{R}%
}+\pi _{\perp }\right\} $ and $\Omega _{4}=\left\{ n:U_{\mathrm{L}}-\pi
_{\perp }<p_{0}<U_{\mathrm{R}}-\pi _{\perp }\right\} $ if $\delta U<2\pi
_{\perp }$. In the ranges{\large \ }$\Omega _{2}${\large \ }and{\large \ }$%
\Omega _{4}${\large \ }we deal with standing waves $\psi _{n}\left( X\right) 
$ completed by linear superpositions of solutions $_{\zeta }\psi _{n}\left(
X\right) $\ and $^{\zeta }\psi _{n}\left( X\right) ${\large \ }with
corresponding longitudinal fluxes that are equal in magnitude for a given $n$%
. The range $\Omega _{3}$ is nontrivial only for critical steps and
perpendicular momenta $\mathbf{p}_{\perp }$ restricted by the inequality $%
2\pi _{\perp }\leq \delta U$. This range is characterized by bounded
energies, $\Omega _{3}=\left\{ n:U_{\mathrm{L}}+\pi _{\perp }\leq p_{0}\leq
U_{\mathrm{R}}-\pi _{\perp }\right\} $. For noncritical steps $\delta
U<\delta U_{c}$, the range $\Omega _{3}$ is absent.

Stationary plane waves, $_{\zeta }\psi _{n}\left( X\right) $ and $^{\zeta
}\psi _{n}\left( X\right) $, are subjected to the following orthonormality
conditions on the $x=\mathrm{const}$ hyperplane: \emph{\ }%
\begin{eqnarray}
&&\left( \ _{\zeta }\psi _{n},\ _{\zeta ^{\prime }}\psi _{n^{\prime
}}\right) _{x}=\zeta \eta _{\mathrm{L}}\delta _{\zeta ,\zeta ^{\prime
}}\delta _{n,n^{\prime }},\ \ \left( \ ^{\zeta }\psi _{n},\ ^{\zeta ^{\prime
}}\psi _{n^{\prime }}\right) _{x}=\zeta \eta _{\mathrm{R}}\delta _{\zeta
,\zeta ^{\prime }}\delta _{n,n^{\prime }};  \notag \\
&&\left( \psi ,\psi ^{\prime }\right) _{x}=\int \psi ^{\dag }\left( X\right)
\gamma ^{0}\gamma ^{1}\psi ^{\prime }\left( X\right) dtd\mathbf{r}_{\bot }\ ,
\label{c3}
\end{eqnarray}%
where $\eta _{\mathrm{L/R}}=\mathrm{sgn\ }\pi _{0}\left( \mathrm{L/R}\right) 
$ is sign of $\pi _{0}\left( \mathrm{L/R}\right) $. We consider our theory
in a large space-time box that has a spatial volume $V_{\bot
}=\prod\limits_{j=2}^{D}K_{j}$ and the time dimension $T$, where all $K_{j}$
and $T$ are macroscopically large. It is supposed that all the solutions $%
\psi \left( X\right) $ are periodic under transitions from one box to
another. The integration over the transverse coordinates is fulfilled from $%
-K_{j}/2$ to $+K_{j}/2$, and over the time $t$ from $-T/2$ to $+T/2$. Under
these suppositions, one can verify, integrating by parts, that the inner
product (\ref{c3}) does not depend on $x$. We assume that the macroscopic
time{\large \ }$T$ is the system surveillance time.

Solutions (\ref{2.15}) with the asymptotic conditions (\ref{2.16}) have the
following form:%
\begin{eqnarray}
&&\ _{+}\varphi _{n}\left( x\right) =\ _{+}\mathcal{N}U_{n,1}\left(
1+z\right) ^{i\alpha _{1}}\left( 1-z\right) ^{i\alpha _{2}}\hat{M}%
_{n}w_{n,1}\left( \frac{z+1}{2}\right) ,\   \notag \\
&&\ _{-}\varphi _{n}\left( x\right) =\ _{-}\mathcal{N}U_{n,2}\left(
1+z\right) ^{i\alpha _{1}}\left( 1-z\right) ^{i\alpha _{2}}\hat{M}%
_{n}w_{n,2}\left( \frac{z+1}{2}\right) ,  \notag \\
&&\ ^{+}\varphi _{n}\left( x\right) =\ ^{+}\mathcal{N}U_{n,3}\left(
1+z\right) ^{i\alpha _{1}}\left( 1-z\right) ^{i\alpha _{2}}\hat{M}%
_{n}w_{n,3}\left( \frac{z+1}{2}\right) ,  \notag \\
&&\ ^{-}\varphi _{n}\left( x\right) =\ ^{-}\mathcal{N}U_{n,4}\left(
1+z\right) ^{i\alpha _{1}}\left( 1-z\right) ^{i\alpha _{2}}\hat{M}%
_{n}w_{n,4}\left( \frac{z+1}{2}\right) ,  \label{2.18}
\end{eqnarray}%
where the constants $U_{n,i}$, $i=1,2,3,4$, and the normalization constants $%
_{\zeta }\mathcal{N}$ and $\ ^{\zeta }\mathcal{N}$ are:%
\begin{eqnarray}
&&U_{n,1}=\frac{2^{2-i(\alpha _{1}-\alpha _{2})}\alpha _{2}}{a-(\alpha
_{1}-\alpha _{2}-2\chi eE_{0}\sigma ^{2})}e^{\pi \alpha _{2}}\ ,  \notag \\
&&U_{n,2}=\frac{2^{1-i(\alpha _{1}+3\alpha _{2})}b(ia-1)}{(2i\alpha _{2}-1)%
\left[ b-(\alpha _{1}-\alpha _{2}+2\chi eE_{0}\sigma ^{2})\right] }e^{-\pi
\alpha _{2}}\ ,  \notag \\
&&U_{n,3}=\frac{2^{1-ia}b\left( b-a\right) }{a\left( \alpha _{1}-\alpha
_{2}+2\chi eE_{0}\sigma ^{2}\right) +b\left( \alpha _{1}-\alpha _{2}-2\chi
eE_{0}\sigma ^{2}\right) }e^{\pi (\alpha _{2}-a)}\ ,  \notag \\
&&U_{n,4}=\frac{2^{-ib}(1-ia)}{1+i(b-a)}e^{\pi (\alpha _{2}-b)}\ ;
\label{2.19b} \\
&&\ _{\zeta }\mathcal{N}=\, {}_{\zeta }CY,\;\;\ ^{\zeta }\mathcal{N}=\
^{\zeta }CY,\;\;Y=\left( V_{\bot }T\right) ^{-1/2}\ ,  \notag \\
&&\ _{\zeta }C=\left[ 2\left\vert p^{\mathrm{L}}\right\vert \left\vert \pi
_{0}\left( \mathrm{L}\right) -\chi p^{\mathrm{L}}\right\vert \right]
^{-1/2},\;\ ^{\zeta }C=\left[ 2\left\vert p^{\mathrm{R}}\right\vert
\left\vert \pi _{0}\left( \mathrm{R}\right) -\chi p^{\mathrm{R}}\right\vert %
\right] ^{-1/2}\ .  \label{e8}
\end{eqnarray}

Stationary plane waves in the ranges $\Omega _{k}$,\ $k=1,2,4,5$ are usually
used in the potential scattering theory. In each of these ranges sign$\eta _{%
\mathrm{L}}$ and sign$\eta _{\mathrm{R}}$ coincide ($\eta _{\mathrm{L/R}}=1$
for particles and $\eta _{\mathrm{L/R}}=-1$ for antiparticles). We stress
that definitions of particle and antiparticle in the framework of
one-particle quantum theory and QFT are in agreement. In the ranges $\Omega
_{1}${\large \ }and $\Omega _{2}$\ there exist only states of particles
whereas in the ranges $\Omega _{4}$\ and $\Omega _{5}$\ there exist only
states of antiparticles.{\large \ }In these ranges particles and
antiparticles are subjected to the scattering and the reflection only. In
fact, $\psi _{n}\left( X\right) $\ for $m\in \Omega _{2}$ are wave functions
that describe an unbounded motion of particles (electrons) in $x\rightarrow
-\infty $ direction while $\psi _{n}\left( X\right) $ for $n\in \Omega _{4}$
are wave functions that describe an unbounded motion of antiparticles
(positrons) toward $x=+\infty $. Such one-particle interpretation does not
exist in the range $\Omega _{3}$, where sign$\eta _{\mathrm{L}}$ is opposite
to sign$\eta _{\mathrm{R}}$, here one must take a many-particle QFT
consideration into account, in particular, the vacuum instability, see
Appendix \ref{Ap} for details. Note that the range $\Omega _{3}$\ is often
referred to as the Klein zone and the pair creation from the vacuum occurs
in this range, whereas the vacuum is stable in the ranges $\Omega _{k}$,\ $%
k=1,2,4,5$.

It was demonstrated in Ref. \cite{GavGi16} (see sections V and VII and
Appendices C1 and C2) by using one-particle mean currents and the energy
fluxes that the plane waves $\ _{\zeta }\psi _{n}\left( X\right) $\ and $\
^{\zeta }\psi _{n}\left( X\right) ${\large \ }are unambiguously identified
as components of initial and final wave packets of particles and
antiparticles, 
\begin{eqnarray}
&&\mathrm{in-solutions:\ }_{+}\psi _{n_{1}},\ ^{-}\psi _{n_{1}};\mathrm{\ }%
_{-}\psi _{n_{5}},\ ^{+}\psi _{n_{5}};\ \ _{-}\psi _{n_{3}},\ ^{-}\psi
_{n_{3}}\ ;  \notag \\
&&\mathrm{out-solutions:\ }_{-}\psi _{n_{1}},\ ^{+}\psi _{n_{1}};\ _{+}\psi
_{n_{5}},\ ^{-}\psi _{n_{5}};\ \ _{+}\psi _{n_{3}},\ ^{+}\psi
_{n_{3}},\;n_{k}\in \Omega _{k}.  \label{in-out}
\end{eqnarray}%
In the ranges $\Omega _{2}$ and $\Omega _{4}$ we deal with a total
reflection. The complete sets of \textrm{in}- and \textrm{out}-solutions
must include solutions $\psi _{n_{2}}\left( X\right) $ and $\psi
_{n_{4}}\left( X\right) $.

Since each pair of solutions $\ _{\zeta }\psi _{n}\left( X\right) $ and $\
^{\zeta }\psi _{n}\left( X\right) $ with quantum numbers $n\in \Omega
_{1}\cup \Omega _{3}\cup \Omega _{5}$ are complete, there exist mutual
decompositions:%
\begin{eqnarray}
\eta _{\mathrm{L}}\ ^{\zeta }\psi _{n}\left( X\right) &=&\ _{+}\psi
_{n}\left( X\right) g\left( _{+}|^{\zeta }\right) -\ _{-}\psi _{n}\left(
X\right) g\left( _{-}|^{\zeta }\right) \,,  \notag \\
\eta _{\mathrm{R}}\ _{\zeta }\psi _{n}\left( X\right) &=&\ ^{+}\psi
_{n}\left( X\right) g\left( ^{+}|_{\zeta }\right) -\ ^{-}\psi _{n}\left(
X\right) g\left( ^{-}|_{\zeta }\right) \,,  \label{2.25}
\end{eqnarray}%
where the decomposition coefficients $g$ are:%
\begin{equation}
g\left( ^{\zeta ^{\prime }}|_{\zeta }\right) ^{\ast }=g\left( _{\zeta
}|^{\zeta ^{\prime }}\right) =\left( \ _{\zeta }\psi _{n},\ ^{\zeta ^{\prime
}}\psi _{n}\right) _{x}\,,\ \ n\in \Omega _{1}\cup \Omega _{3}\cup \Omega
_{5}\,.  \label{2.26}
\end{equation}%
These coefficients satisfy the following unitary relations:%
\begin{eqnarray}
&&\left\vert g\left( _{-}\left\vert ^{+}\right. \right) \right\vert
^{2}=\left\vert g\left( _{+}\left\vert ^{-}\right. \right) \right\vert
^{2},\;\left\vert g\left( _{+}\left\vert ^{+}\right. \right) \right\vert
^{2}=\left\vert g\left( _{-}\left\vert ^{-}\right. \right) \right\vert ^{2},
\notag \\
&&\frac{g\left( _{+}\left\vert ^{-}\right. \right) }{g\left( _{-}\left\vert
^{-}\right. \right) }=\frac{g\left( ^{+}\left\vert _{-}\right. \right) }{%
g\left( ^{+}\left\vert _{+}\right. \right) },\ \left\vert g\left(
_{+}\left\vert ^{-}\right. \right) \right\vert ^{2}-\left\vert g\left(
_{+}\left\vert ^{+}\right. \right) \right\vert ^{2}=-\eta _{\mathrm{L}}\eta
_{\mathrm{R}}\ .  \label{UR}
\end{eqnarray}

One can see that all the coefficients $g$ can be expressed via only one of
them, e.g. via $g\left( ^{-}|_{+}\right) $. Using the Kummer relations (\ref%
{cm1}) for the hypergeometric equation; see \cite{Bateman}, this coefficient
can be found to be:%
\begin{equation}
g\left( ^{-}|_{+}\right) =i\eta _{\mathrm{R}}\,\frac{_{+}\mathcal{N}}{^{-}%
\mathcal{N}}\frac{2^{i(b-\alpha _{1}+\alpha _{2})+1}\Gamma \left[ i(a-b)%
\right] \Gamma \left( 1+2i\alpha _{2}\right) }{\left( a-\alpha _{1}+\alpha
_{2}+2\chi eE_{0}\sigma ^{2}\right) \Gamma \left( ia\right) \Gamma \left[
-i(b-2\alpha _{2})\right] },  \label{2.20}
\end{equation}%
where $\Gamma \left( x\right) $ is the gamma-function. Then%
\begin{eqnarray}
&&\left\vert g\left( _{+}\left\vert ^{-}\right. \right) \right\vert
^{-2}=\sinh \left( 2\pi \left\vert p^{\mathrm{L}}\right\vert \sigma \right)
\sinh \left( 4\pi \left\vert p^{\mathrm{R}}\right\vert \sigma \right)
\left\vert \beta _{+}\beta _{-}\right\vert ^{-1}\ ,  \notag \\
&&\beta _{\pm }=\sinh \left[ \pi \sigma \left( \sqrt{2\delta
U^{2}+2\left\vert p^{\mathrm{R}}\right\vert ^{2}-\left\vert p^{\mathrm{L}%
}\right\vert ^{2}}\pm \left( 2\left\vert p^{\mathrm{R}}\right\vert
-\left\vert p^{\mathrm{L}}\right\vert \right) \right) \right] ,  \label{Ncr}
\end{eqnarray}%
where $\delta U$ and $\left\vert p^{\mathrm{L/R}}\right\vert $ are given by
Eqs. \ref{dU} and \ref{2.12}).

Relation (\ref{Ncr}) holds true for the quantum numbers $n\in \Omega
_{1}\cup \Omega _{3}\cup \Omega _{5}$.\ However, interpretations of this
relation in the range $\Omega _{3}$ and in the ranges $\Omega _{1}$ and $%
\Omega _{5}$ are quite different. Note that there exists a useful relation
between absolute values of the momenta $p^{\mathrm{R}}$ and $p^{\mathrm{L}},$%
\begin{eqnarray}
\left\vert p^{\mathrm{L}}\right\vert &=&\sqrt{\left\vert p^{\mathrm{R}%
}\right\vert ^{2}+2\eta _{\mathrm{R}}\delta U\sqrt{\left\vert p^{\mathrm{R}%
}\right\vert ^{2}+\pi _{\bot }^{2}}+\delta U^{2}},  \notag \\
\left\vert p^{\mathrm{R}}\right\vert &=&\sqrt{\left\vert p^{\mathrm{L}%
}\right\vert ^{2}-2\eta _{\mathrm{L}}\delta U\sqrt{\left\vert p^{\mathrm{L}%
}\right\vert ^{2}+\pi _{\bot }^{2}}+\delta U^{2}},  \label{pLR}
\end{eqnarray}%
As follows from Eqs.~(\ref{Ncr}) and (\ref{pLR}), if either $\left\vert p^{%
\mathrm{R}}\right\vert $ or $\left\vert p^{\mathrm{L}}\right\vert $ tends to
zero, one of the following limits takes place:%
\begin{equation}
\left\vert g\left( _{-}\left\vert ^{+}\right. \right) \right\vert ^{-2}\sim
\left\vert p^{\mathrm{R}}\right\vert \rightarrow 0,\ \ \left\vert g\left(
_{-}\left\vert ^{+}\right. \right) \right\vert ^{-2}\sim \left\vert p^{%
\mathrm{L}}\right\vert \rightarrow 0,\ \ \forall \pi _{\bot }^{2}\neq 0.
\label{L15}
\end{equation}%
These properties are essential for the justification of \textrm{in}- and 
\textrm{out}-particle interpretation in the general construction given in
Ref. \cite{GavGi16}.

\section{Processes in stable vacuum ranges\label{S3}}

In the ranges $\Omega _{2}$ and $\Omega _{4}$ we deal with a total
reflection. In the adjacent ranges,\emph{\ }$\Omega _{1}$ and $\Omega _{5}$,
a particle can be reflected and transmitted. For example, in the range $%
\Omega _{1},$ the total $\tilde{R}$ and the relative $R$ amplitudes of an
electron reflection, and the total $\tilde{T}$\ and the relative $T$
amplitudes of an electron transmission can be presented via the following
matrix elements:%
\begin{eqnarray}
R_{+,n} &=&\tilde{R}_{+,n}c_{v}^{-1},\ \tilde{R}_{+,n}=\langle 0,\mathrm{out}%
|\ _{-}a_{n}(\mathrm{out})\ _{+}a_{n}^{\dag }(\mathrm{in})|0,\mathrm{in}%
\rangle ,  \notag \\
T_{+,n} &=&\tilde{T}_{+,n}c_{v}^{-1},\ \tilde{T}_{+,n}=\langle 0,\mathrm{out}%
|\ ^{+}a_{n}(\mathrm{out})\ _{+}a_{n}^{\dag }(\mathrm{in})|0,\mathrm{in}%
\rangle ,  \notag \\
R_{-,n} &=&\tilde{R}_{-,n}c_{v}^{-1},\ \tilde{R}_{-,n}=\langle 0,\mathrm{out}%
|\ ^{+}a_{n}(\mathrm{out})\ ^{-}a_{n}^{\dag }(\mathrm{in})|0,\mathrm{in}%
\rangle ,  \notag \\
T_{-,n} &=&\tilde{T}_{-,n}c_{v}^{-1},\ \tilde{T}_{-,n}=\langle 0,\mathrm{out}%
|_{-}a_{n}(\mathrm{out})\ ^{-}a_{n}^{\dag }(\mathrm{in})|0,\mathrm{in}%
\rangle ,  \label{L13}
\end{eqnarray}%
where initial creation $a^{\dag }(\mathrm{in})$ and final annihilation $a(%
\mathrm{out})$ operators, initial $|0,\mathrm{in}\rangle $ and final $|0,%
\mathrm{out}\rangle $ vacua, and the vacuum-to-vacuum transition amplitude $%
c_{v}=\langle 0,\mathrm{out}|0,\mathrm{in}\rangle $ are defined in Appendix %
\ref{Ap}. Note that the partial vacua are stable in $\Omega _{k}$, $%
k=1,2,4,5 $, and the vacuum instability with $\left\vert c_{v}\right\vert
\neq 1$ is due to the partial vacuum-to-vacuum transition amplitude formed
in $\Omega _{3}$. Using a linear canonical transformation between \textrm{in}
and \textrm{out}-operators in Eq. (\ref{L13}) (see Eq. (4.33) in Ref. \cite%
{GavGi16}) one find that the relative reflection $\left\vert R_{\zeta
,n}\right\vert ^{2}$ and transmission $\left\vert T_{\zeta ,n}\right\vert
^{2}$ probabilities are:\textrm{\ }%
\begin{equation}
\left\vert T_{\zeta ,n}\right\vert ^{2}=1-\left\vert R_{\zeta ,n}\right\vert
^{2},\;\;\left\vert R_{\zeta ,n}\right\vert ^{2}=\left[ 1+\left\vert g\left(
_{+}\left\vert ^{-}\right. \right) \right\vert ^{-2}\right] ^{-1},\ \zeta
=\pm \ ,  \label{L14}
\end{equation}%
where $\left\vert g\left( _{+}\left\vert ^{-}\right. \right) \right\vert
^{-2}$ is given by Eq. (\ref{Ncr}). \ Similar expressions can be derived for
positron amplitudes in the range $\Omega _{5}$. In particular, relation (\ref%
{L14}) holds true literally for the positrons in the range $\Omega _{5}$.

In the ranges $\Omega _{1}$\ and $\Omega _{5}$\ we meet a realization of
rules of the potential scattering theory in the framework of QFT and can see
that relative probabilities of the reflection and the transmission coincide
with mean currents of reflected particles $J_{R}=\left\vert R_{m}\right\vert
^{2}$\ and transmitted particles $J_{T}=\left\vert T_{m}\right\vert ^{2}$.
The correct result $J_{R}+J_{T}=1$\ follows from the unitary relation (\ref%
{UR}).

Limits (\ref{L15}) imply the following properties of the coefficients $%
\left\vert g\left( _{+}\left\vert ^{-}\right. \right) \right\vert $: $%
\left\vert g\left( _{+}\left\vert ^{-}\right. \right) \right\vert
^{-2}\rightarrow 0$ in the range $\Omega _{1}$ if $n$ tends to the boundary
with the range $\Omega _{2}$ ($\left\vert p^{\mathrm{R}}\right\vert
\rightarrow 0$); $\left\vert g\left( _{+}\left\vert ^{-}\right. \right)
\right\vert ^{-2}\rightarrow 0$ in the range $\Omega _{5}$ if $n$ tends to
the boundary with the range $\Omega _{4}$ ($\left\vert p^{\mathrm{L}%
}\right\vert \rightarrow 0$). Thus, in the latter cases the relative
reflection probabilities $\left\vert R_{\zeta ,n}\right\vert ^{2}$ tend to
the unity; i.e., they are continuous functions of the quantum numbers $n$ on
the boundaries. In addition, it follows from Eq.~(\ref{2.20}) that $%
\left\vert g\left( _{+}\left\vert ^{-}\right. \right) \right\vert
^{-2}\rightarrow 0$ and, therefore, $\left\vert R_{\zeta ,n}\right\vert
^{2}\rightarrow 0$ as $p_{0}\rightarrow \pm \infty $, as it is expected.

\section{Physical quantities specifying the vacuum instability\label{S4}}

The vacuum instability is due to contributions formed in the range $\Omega
_{3}$. In this range the important characteristic of all the processes are
differential mean numbers $N_{n}^{\mathrm{cr}}$ of electron-positron pairs
created from the vacuum. The differential mean numbers of electrons and
positrons created from the vacuum are equal and related to the mean numbers $%
N_{n}^{\mathrm{cr}}$ of created pairs,%
\begin{align}
& N_{n}^{a}\left( \mathrm{out}\right) =\left\langle 0,\mathrm{in}\left\vert 
\hat{N}_{n}^{a}\left( \mathrm{out}\right) \right\vert 0,\mathrm{in}%
\right\rangle =\left\vert g\left( _{-}\left\vert ^{+}\right. \right)
\right\vert ^{-2}\ ,  \notag \\
& N_{n}^{b}\left( \mathrm{out}\right) =\left\langle 0,\mathrm{in}\left\vert 
\hat{N}_{n}^{b}\left( \mathrm{out}\right) \right\vert 0,\mathrm{in}%
\right\rangle =\left\vert g\left( _{+}\left\vert ^{-}\right. \right)
\right\vert ^{-2}\ ,  \notag \\
& N_{n}^{\mathrm{cr}}=N_{n}^{b}\left( \mathrm{out}\right) =N_{n}^{a}\left( 
\mathrm{out}\right) ,\ \ n\in \Omega _{3}\ .  \label{7.5}
\end{align}%
Here $\hat{N}_{n}^{a}\left( \mathrm{out}\right) $ and $N_{n}^{b}\left( 
\mathrm{out}\right) $ are operators of the number of the final electrons and
positrons, given by Eq. (\ref{Nop}) (see Appendix \ref{Ap} for details), and
the quantity $\left\vert g\left( _{+}\left\vert ^{-}\right. \right)
\right\vert ^{-2}$ is given by Eq. (\ref{2.20}). The probabilities of a
particle reflection (transmission is impossible) and a pair creation and
annihilation in the Klein zone can be expressed via differential mean
numbers of created pairs $N_{n}^{\mathrm{cr}}$; see Eq. (7.22) in Ref.\emph{%
\ \cite{GavGi16}})\emph{.}

Unlike the case of uniform time-dependent electric fields, in the constant
inhomogeneous electric fields, there is a critical surface in space of
particle momenta, which separates the Klein zone $\Omega _{3}$ from the
adjacent ranges $\Omega _{2}$ and $\Omega _{4}$. In the ranges $\Omega _{2}$
and $\Omega _{4}$, the work of the electric field is sufficient to ensure
the total reflection for electrons and positrons, respectively, but is not
sufficient to produce pairs from the vacuum. Accordingly, it is expected
that for any nonpathological field configuration, the pair creation vanishes
close to this critical surface. Limits (\ref{L15}) imply that $N_{n}^{%
\mathrm{cr}}\rightarrow 0$ if $n$ tends to the boundary with either the
range $\Omega _{2}$ ($\left\vert p^{\mathrm{R}}\right\vert \rightarrow 0$)
or the range $\Omega _{4}$ ($\left\vert p^{\mathrm{L}}\right\vert
\rightarrow 0$),%
\begin{equation}
N_{n}^{\mathrm{cr}}\sim \left\vert p^{\mathrm{R}}\right\vert \rightarrow 0,\
\ N_{n}^{\mathrm{cr}}\sim \left\vert p^{\mathrm{L}}\right\vert \rightarrow
0,\ \ \forall \pi _{\bot }\neq 0\ .  \label{Nb}
\end{equation}

Standard integral characteristics of the vacuum instability are sums over
the range $\Omega _{3}$ (see Appendix \ref{Ap} for details), are the total
number $N^{\mathrm{cr}}$ of pairs\ created from the vacuum, and the
vacuum-to-vacuum transition probability $P_{v}$,

\begin{eqnarray}
&&N^{\mathrm{cr}}=\sum_{n\in \Omega _{3}}N_{n}^{\mathrm{cr}},\;N_{n}^{%
\mathrm{cr}}=\left\vert g\left( _{+}\left\vert ^{-}\right. \right)
\right\vert ^{-2},  \label{3.3} \\
&&P_{v}=\left\vert c_{v}\right\vert ^{2}=\exp \left( \sum_{n\in \Omega
_{3}}\ln \left( 1-N_{n}^{\mathrm{cr}}\right) \right) .  \label{3.5}
\end{eqnarray}%
The summations over $\Omega _{3}$ can be converted into integrals in the
standard way, 
\begin{equation*}
\left( V_{\perp }T\right) ^{-1}\sum {}_{p_{0},\mathbf{p}_{\perp }\in \Omega
_{3}}\leftrightarrow \left( 2\pi \right) ^{1-d}\int dp_{0}d\mathbf{p}_{\perp
}\ ,
\end{equation*}%
in which $V_{\perp }$, $T$ are macroscopically large. It follows from Eq. (%
\ref{3.5}) that $\ln P_{v}\approx -N^{\mathrm{cr}}$ if all $N_{n}^{\mathrm{cr%
}}\ll 1$.

Under approximation (\ref{asympt}) the electric field under consideration
can be neglected in the macroscopic regions $S_{\mathrm{L}}$ (at $x<x_{%
\mathrm{L}})$ and $S_{\mathrm{R}}$ (at $x>x_{\mathrm{R}})$, that is,
particles are free in these regions. We note that near all the work $\delta
U $ is performed by the electric field situated in a region{\large \ }$S_{%
\mathrm{int}}${\large \ }between two planes $x=x_{\mathrm{L}}$ and $x=x_{%
\mathrm{R}}$. Assuming that the areas $S_{\mathrm{L}}$\ and $S_{\mathrm{R}}$%
\ are much wider than the area\ $S_{\mathrm{int}}$, this part of the field
affects only coefficients $g$ entering into the mutual decompositions of the
solutions given by Eq. (\ref{2.25}). Created electrons and positrons leaving
the area $S_{\mathrm{int}}$ enter the areas $S_{\mathrm{L}}$ and $S_{\mathrm{%
R}}$, respectively, and continue to move with constant velocities. The
positron of a pair created with quantum number $n$ moves in the $x$
direction with a velocity $v^{\mathrm{R}}=\left\vert p^{\mathrm{R}}/\pi
_{0}\left( \mathrm{R}\right) \right\vert $ while the electron belonging to
the same pair moves in the opposite direction with a velocity $-v^{\mathrm{L}%
}$, $v^{\mathrm{L}}=\left\vert p^{\mathrm{L}}/\pi _{0}\left( \mathrm{L}%
\right) \right\vert $. It is shown that the microscopical parameter $T$ can
be interpreted as the time of the observation of the created particles
leaving the area $S_{\mathrm{int}}$; see \cite{GavGi20}.

Following the way used in Ref. \cite{GavGi20}, we can calculate the current
densities and the energy flux densities of electrons and positrons, after
the instant when these fluxes become completely separated and already have
left the region $S_{\mathrm{int}}$. The motion of the positrons forms the
flux density%
\begin{equation}
\left\langle j_{x}\right\rangle _{n}=N_{n}^{\mathrm{cr}}(TV_{\perp })^{-1}
\label{d19a}
\end{equation}%
in the area $S_{\mathrm{R}}$, while the electron motion forms the flux
density $-\left( j_{x}\right) _{n}$ in the area $S_{\mathrm{L}}$. Here it is
taken into account that differential mean numbers of created electrons and
positrons with a given $n$ are equal. The total flux densities of the
positrons and electrons are%
\begin{equation}
\left\langle j_{x}\right\rangle =\sum_{n\in \Omega _{3}}\left\langle
j_{x}\right\rangle _{n}=N^{\mathrm{cr}}(TV_{\perp })^{-1}  \label{d19b}
\end{equation}%
and $-\left\langle j_{x}\right\rangle $, respectively. The current density
of both created electrons and positrons is $J_{x}^{\mathrm{cr}%
}=e\left\langle j_{x}\right\rangle $. It is conserved in the $x$-direction.

During the time $T,$ the created positrons carry the charge $e\left\langle
j_{x}\right\rangle _{n}T$ over the unit area $V_{\bot }$ of the surface $%
x=x_{\mathrm{R}}$. This charge is evenly distributed over the cylindrical
volume of the length $v^{\mathrm{R}}T$. Thus, the charge density of the
positrons created with a given $n$ is $ej_{n}^{0}\left( \mathrm{R}\right) $,
where $j_{n}^{0}\left( \mathrm{R}\right) =\left\langle j_{x}\right\rangle
_{n}/v^{\mathrm{R}}$ is the number density of the positrons. During the time 
$T,$ the created electrons carry the charge $e\left\langle
j_{x}\right\rangle _{n}T$ over the unit area $V_{\bot }$ of the surface $%
x=x_{\mathrm{L}}$. Taking into account that this charge is evenly
distributed over the cylindrical volume of the length $v^{\mathrm{L}}T$, we
can see that the charge density of the electrons created with a given $n$ is 
$-ej_{n}^{0}\left( \mathrm{L}\right) $, where $j_{n}^{0}\left( \mathrm{L}%
\right) =\left\langle j_{x}\right\rangle _{n}/v^{\mathrm{L}}$ is the number
density of the electrons. The total charge density of the created particles
reads:%
\begin{equation}
J_{0}^{\mathrm{cr}}\left( x\right) =e\left\{ 
\begin{array}{c}
-\sum_{n\in \Omega _{3}}j_{n}^{0}\left( \mathrm{L}\right) ,\ \ x\in S_{%
\mathrm{L}} \\ 
\sum_{n\in \Omega _{3}}j_{n}^{0}\left( \mathrm{R}\right) ,\ x\in S_{\mathrm{R%
}}%
\end{array}%
\right. .\   \label{d22}
\end{equation}%
Due to a relation between the velocities $v^{\mathrm{L}}$ and $v^{\mathrm{R}%
} $, the total number densities of the created electrons and positrons are
the same,%
\begin{equation*}
\sum_{n\in \Omega _{3}}j_{n}^{0}\left( \mathrm{L}\right) =\sum_{n\in \Omega
_{3}}j_{n}^{0}\left( \mathrm{R}\right) \ .
\end{equation*}%
The created electrons and positrons are spatially separated and carry a
charge that tends to weaken the external electric field.

In the same manner, one can derive some representation for the nonzero
components of energy-momentum tensor of the created particles:%
\begin{eqnarray}
&&T_{\mathrm{cr}}^{00}(x)=\left\{ 
\begin{array}{c}
\sum_{n\in \Omega _{3}}j_{n}^{0}\left( \mathrm{L}\right) \pi _{0}\left( 
\mathrm{L}\right) ,\ x\in S_{\mathrm{L}} \\ 
\sum_{n\in \Omega _{3}}j_{n}^{0}\left( \mathrm{R}\right) \left\vert \pi
_{0}\left( \mathrm{R}\right) \right\vert ,\mathrm{\ }x\in S_{\mathrm{R}}%
\end{array}%
\right. ,  \notag \\
&&T_{\mathrm{cr}}^{11}(x)=\left\{ 
\begin{array}{c}
\sum_{n\in \Omega _{3}}\left\langle j_{x}\right\rangle _{n}\left\vert p^{%
\mathrm{L}}\right\vert ,\mathrm{\ }x\in S_{\mathrm{L}} \\ 
\sum_{n\in \Omega _{3}}\left\langle j_{x}\right\rangle _{n}\left\vert p^{%
\mathrm{R}}\right\vert ,\mathrm{\ }x\in S_{\mathrm{R}}%
\end{array}%
\right. ,  \notag \\
&&T_{\mathrm{cr}}^{kk}(x)=\left\{ 
\begin{array}{c}
\sum_{n\in \Omega _{3}}\left\langle j_{x}\right\rangle _{n}\left(
p_{k}\right) ^{2}/\left\vert p^{\mathrm{L}}\right\vert ,\mathrm{\ }x\in S_{%
\mathrm{L}} \\ 
\sum_{n\in \Omega _{3}}\left\langle j_{x}\right\rangle _{n}\left(
p_{k}\right) ^{2}/\left\vert p^{\mathrm{R}}\right\vert ,\mathrm{\ }x\in S_{%
\mathrm{R}}%
\end{array}%
\right. ,\ \ k\neq 1,  \notag \\
&&T_{\mathrm{cr}}^{10}(x)=\left\{ 
\begin{array}{c}
-\sum_{n\in \Omega _{3}}\left\langle j_{x}\right\rangle _{n}\pi _{0}\left( 
\mathrm{L}\right) ,\mathrm{\ }x\in S_{\mathrm{L}} \\ 
\sum_{n\in \Omega _{3}}\left\langle j_{x}\right\rangle _{n}\left\vert \pi
_{0}\left( \mathrm{R}\right) \right\vert ,\mathrm{\ }x\in S_{\mathrm{R}}%
\end{array}%
\right. .  \label{d23}
\end{eqnarray}%
Here $T_{\mathrm{cr}}^{00}(x)$ and $T_{\mathrm{cr}}^{kk}(x)$, $k=1,\ldots ,D$%
, (there is no summation over $k$) are energy density and components of the
pressure of the particles created in the areas $S_{\mathrm{L}}$ and $S_{%
\mathrm{R}}$ respectively, whereas $T_{\mathrm{cr}}^{10}(x)v_{s}$, for $x\in
S_{\mathrm{L}}$ or $x\in S_{\mathrm{R}}$, is the energy flux density of the
created particles\ through the surfaces $x=x_{\mathrm{L}}$ or $x=x_{\mathrm{R%
}}$ respectively. In a strong field, or in a field with the sufficiently
large potential step $\delta U$ , the energy density and the pressure{\large %
\ }along the direction of the axis $x$ are near equal, $T_{\mathrm{cr}%
}^{00}(x)\approx T_{\mathrm{cr}}^{11}(x)$, in the areas $S_{\mathrm{L}}$%
\emph{\ }and\emph{\ }$S_{\mathrm{R}}$ respectively,\emph{\ }

\section{Particle creation due to a weakly inhomogeneous electric field\label%
{S5}}

\subsection{Intensity of the particle creation over the Klein zone\label{5.1}%
}

The above study of the vacuum instability caused by the asymmetric $x$-step
can be useful for a consideration of the particle creation by a weakly
inhomogeneous electric field between two capacitor plates separated by a
sufficiently large length. Indeed, if the parameter $\sigma $ is taken to be
sufficiently large,%
\begin{equation}
\sigma \gg \left( eE_{0}\right) ^{-1/2}\max \left\{ 1,m^{2}/eE_{0}\right\} ,
\label{3.12}
\end{equation}%
the step can be considered as a regularization (like the Sauter potential
with appropriate parameters) of a weakly inhomogeneous constant electric
field between the plates. For example, for such big $\sigma $ we can
consider the behavior of mean numbers of electron-positron pairs created
over the Klein zone\emph{\ }$\Omega _{3}$. For this purpose, consider
arguments of the functions $\beta _{\pm }$ in the denominator of expression (%
\ref{Ncr}). Absolute values of $\left\vert p^{\mathrm{R}}\right\vert $ and $%
\left\vert p^{\mathrm{L}}\right\vert $ are related by Eq. (\ref{pLR}). One
can see from Eq. (\ref{pLR}) that $d\left\vert p^{\mathrm{L}}\right\vert
/d\left\vert p^{\mathrm{R}}\right\vert <0,$ and at any given $\mathbf{p}%
_{\bot }$ these quantities are restricted inside the range $\Omega _{3}$ as%
\begin{equation}
0\leq \left\vert p^{\mathrm{R/L}}\right\vert \leq p^{\mathrm{\max }},\;\;p^{%
\mathrm{\max }}=\sqrt{\delta U\left( \delta U-2\pi _{\bot }\right) }.
\label{d8}
\end{equation}%
These relations for big $\sigma $ are%
\begin{equation}
\pi \sigma \left( \sqrt{\delta U\left( \delta U\mathbb{-\pi }_{\perp
}\right) }-p^{\mathrm{\max }}\right) \gg 1,\ \ \pi \sigma \left( \sqrt{%
\delta U\left( \delta U\mathbb{+}2\mathbb{\pi }_{\perp }\right) }-p^{\mathrm{%
\max }}\right) \gg 1,  \label{s2}
\end{equation}%
and imply that 
\begin{equation}
\pi \sigma \left[ \sqrt{2\delta U^{2}+2\left\vert p^{\mathrm{R}}\right\vert
^{2}-\left\vert p^{\mathrm{L}}\right\vert ^{2}}\pm \left( 2\left\vert p^{%
\mathrm{R}}\right\vert -\left\vert p^{\mathrm{L}}\right\vert \right) \right]
\gg 1.  \label{s3}
\end{equation}%
We get from (\ref{Ncr}) that%
\begin{equation}
N_{n}^{\mathrm{cr}}=\left\vert g\left( _{+}\left\vert ^{-}\right. \right)
\right\vert ^{-2}\approx \frac{4\sinh \left( 2\pi \left\vert p^{\mathrm{L}%
}\right\vert \sigma \right) \sinh \left( 4\pi \left\vert p^{\mathrm{R}%
}\right\vert \sigma \right) }{\exp \left[ 2\pi \sigma \sqrt{2\delta
U^{2}+2\left\vert p^{\mathrm{R}}\right\vert ^{2}-\left\vert p^{\mathrm{L}%
}\right\vert ^{2}}\right] }\ .  \label{a1}
\end{equation}%
It follows from Eq. (\ref{a1}) that the quantities $N_{n}^{\mathrm{cr}}$\
are exponentially small,%
\begin{equation}
N_{n}^{\mathrm{cr}}\approx 2\left( 4\pi \sigma \right) ^{2}\left\vert p^{%
\mathrm{L}}p^{\mathrm{R}}\right\vert \exp \left( -2\sqrt{2}\pi \delta
U\sigma \right) \ ,  \label{n4}
\end{equation}%
if the range $\Omega _{3}$ is small enough 
\begin{equation}
eE\sigma -2\pi _{\bot }\rightarrow 0\Longrightarrow \pi \sigma p^{\mathrm{%
\max }}\ll 1\Longrightarrow \left\vert p^{\mathrm{R/L}}\right\vert \ll 1\ .
\label{exs10}
\end{equation}

Then we can consider the opposite case of big ranges $\Omega _{3}$ when%
\begin{equation}
\pi \sigma p^{\mathrm{\max }}\gg 1  \label{exs11}
\end{equation}%
and the quantities $N_{n}^{\mathrm{cr}}$ are not small. Such ranges do exist
if%
\begin{equation}
eE\sigma \gg m  \label{exs12}
\end{equation}%
and 
\begin{equation}
\sigma \pi _{\bot }<K_{\bot }\ ,  \label{exs11b}
\end{equation}%
where $K_{\bot }$ is a given arbitrary number, restricted as $m\sigma \ll
K_{\bot }\ll eE\sigma ^{2}$.

Let us study the behavior of $N_{n}^{\mathrm{cr}}$ on the boundaries of the
Klein domain $\Omega _{3}$, when $\left\vert p^{\mathrm{R}}\right\vert
\rightarrow 0$ or $\left\vert p^{\mathrm{L}}\right\vert \rightarrow 0$. Let $%
\pi \sigma \left\vert p^{\mathrm{R/L}}\right\vert <K_{1}$, where $K_{1}\geq
1 $ is some arbitrary number satisfying the inequality%
\begin{equation}
K_{1}\ll eE_{0}\sigma ^{2}\ .  \label{e9}
\end{equation}

In close proximity to these boundaries, $\pi \sigma \left\vert p^{\mathrm{L/R%
}}\right\vert <K_{0}$, $K_{0}<1$, we obtain that the value $N_{n}^{\mathrm{cr%
}}$ is exponentially small,%
\begin{equation}
N_{n}^{\mathrm{cr}}\approx 2e^{-4\pi m\sigma },\ \ \pi \sigma \left\vert p^{%
\mathrm{R}}\right\vert <K_{0}\ ;\ \ N_{n}^{\mathrm{cr}}\approx 2e^{-2m\sigma
},\ \ \pi \sigma \left\vert p^{\mathrm{L}}\right\vert <K_{0}\ .  \label{i3d}
\end{equation}%
For boundary regions located closer to the center of the Klein zone $\Omega
_{3}$, $1\lesssim \pi \sigma \left\vert p^{\mathrm{L/R}}\right\vert <K_{1}$, 
$K_{1}>1$ the following approximation is valid%
\begin{eqnarray}
N_{n}^{\mathrm{cr}} &\approx &\exp \left\{ -4\pi \sigma \left[ \sqrt{\left(
p^{\mathrm{R}}\right) ^{2}+\pi _{\bot }^{2}}-\left\vert p^{\mathrm{R}%
}\right\vert \right] \right\} ,\ \ 1\lesssim \pi \sigma \left\vert p^{%
\mathrm{R}}\right\vert \lesssim K_{1}\ ,  \notag \\
N_{n}^{\mathrm{cr}} &\approx &\exp \left\{ -2\pi \sigma \left[ \sqrt{\left(
p^{\mathrm{L}}\right) ^{2}+\pi _{\bot }^{2}}-\left\vert p^{\mathrm{L}%
}\right\vert \right] \right\} ,\ \ 1\lesssim \pi \sigma \left\vert p^{%
\mathrm{L}}\right\vert \lesssim K_{1}\ .  \label{exs16}
\end{eqnarray}

The numbers $N_{n}^{\mathrm{cr}}$ increase as $n$ moves away from the
boundaries of the Klein region $\Omega _{3}$ and for any fixed value of $\pi
_{\bot }$, they reach their maximum value when $\pi \sigma \left\vert p^{%
\mathrm{L/R}}\right\vert \rightarrow K_{1}$. In turn, this maximum value
increases as $\pi _{\bot }\rightarrow m$. In this range we can estimate $%
N_{n}^{\mathrm{cr}}$ as%
\begin{equation}
N_{n}^{\mathrm{cr}}<\exp \left\{ -4\pi \sigma m\left[ \frac{\pi \sigma m}{%
2K_{1}}-\left( \frac{\pi \sigma m}{2K_{1}}\right) ^{3}\right] \right\} ,\ \
\pi \sigma \left\vert p^{\mathrm{R}}\right\vert \rightarrow K_{1}\ ,
\label{i4}
\end{equation}%
Similarly, in the domain $\pi \sigma \left\vert p^{\mathrm{L}}\right\vert
\rightarrow k$ we have the estimate%
\begin{equation}
N_{n}^{\mathrm{cr}}<\exp \left\{ -2\pi \sigma m\left[ \frac{\pi \sigma m}{%
2K_{1}}-\left( \frac{\pi \sigma m}{2K_{1}}\right) ^{3}\right] \right\} ,\ \
\pi \sigma \left\vert p^{\mathrm{L}}\right\vert \rightarrow K_{1}\ .
\label{i5}
\end{equation}%
The right sides of inequalities (\ref{i4}) and (\ref{i5}) are exponentially
small if 
\begin{equation}
K_{1}\ll \pi m\sigma /2  \label{exs31}
\end{equation}%
for any $K_{1}$.

Consequently, the main contribution to the pairs creation is formed in the
subrange of the Klein zone $D\subset \Omega _{3}$, where the energy $\pi
_{\bot }$ is restricted by inequality (\ref{exs11b}) and the energy $p_{0}$
is restricted as: 
\begin{equation}
-eE_{0}\sigma ^{2}+K<\sigma p_{0}<-K,\ \ \,K=\sqrt{K_{1}^{2}+(\sigma \pi
_{\bot })^{2}}\ .  \label{exs21}
\end{equation}%
Inequalities (\ref{exs11b}) and (\ref{e9}) imply that $K\ll eE\sigma ^{2}$.
In such a case, we can approximate numbers (\ref{a1}) as follows%
\begin{equation}
N_{n}^{\mathrm{cr}}\approx N_{p_{0},\mathbf{p}_{\bot }}^{\mathrm{as}%
}=e^{-\pi \tau },\ \,\tau =2\sigma \left( \sqrt{2\delta U^{2}+2\left\vert p^{%
\mathrm{R}}\right\vert ^{2}-\left\vert p^{\mathrm{L}}\right\vert ^{2}}%
-\left\vert p^{\mathrm{L}}\right\vert -2\left\vert p^{\mathrm{R}}\right\vert
\right) \ .  \label{exs22}
\end{equation}

In the point $x_{\mathrm{M}}=\sigma \ln 2$ the electric field (\ref{1.2})
has the maximum value $E_{\max }=E_{0}/(3\sqrt{3})$ and the corresponding
kinetic energy reads:%
\begin{equation*}
\pi _{0}^{\prime }=p_{0}-U\left( x_{\mathrm{M}}\right) =p_{0}+\frac{%
eE_{0}\sigma }{\sqrt{3}}\ .
\end{equation*}%
The function $\tau $ takes its minimum value at the point $\pi _{0}^{\prime
}=0$,%
\begin{equation}
\min \tau =\left. \tau \right\vert _{\pi _{0}^{\prime }=0}=\lambda =\frac{%
\pi _{\bot }^{2}}{eE_{\max }}\ .  \label{exs23}
\end{equation}%
Further, $\tau $ monotonically increases with $\left\vert \pi _{0}^{\prime
}\right\vert $ approaching the boundary of the subrange $D$. On the one side
of the center of the Klein zone $\Omega _{3}$, the function $\tau $ takes
the maximum value%
\begin{equation*}
\tau _{\max }^{-}=\left. \tau \right\vert _{\sigma \left\vert
p_{0}\right\vert \rightarrow eE_{0}\sigma ^{2}-K}\simeq 2\left(
K-K_{1}/\sigma +\frac{\lambda }{4\sqrt{3}}\right) ,\ 
\end{equation*}%
while on the other side, the maximum value is:%
\begin{equation*}
\tau _{\max }^{+}=\left. \tau \right\vert _{\sigma \left\vert
p_{0}\right\vert \rightarrow K}\simeq 4\left( K-K_{1}/\sigma \right) \ .
\end{equation*}%
In the wide range $D$, where%
\begin{equation}
K/\sigma -eE_{0}\sigma \left( 1-1/\sqrt{3}\right) <\pi _{0}^{\prime
}<eE_{0}\sigma /\sqrt{3}-K/\sigma \ ,\   \label{exs24}
\end{equation}%
the numbers $N_{n}^{\mathrm{cr}}$ practically do not depend on the parameter 
$\sigma $ and have the form of the differential number of created particles
in an uniform electric field \cite{Nikis70b,Nikis79},%
\begin{equation}
N_{n}^{\mathrm{cr}}\approx e^{-\pi \lambda }\ .  \label{exs23b}
\end{equation}

\subsection{Integral quantities\label{S5.2}}

The total number of\ created pairs is given by integral (\ref{3.3}). The
main contribution to the integral is due to the subrange $D\subset \Omega
_{3}$, defined by Eqs.~(\ref{exs11b}) and (\ref{exs21}). In this subrange
the functions $N_{n}^{\mathrm{cr}}$ can be approximated by Eq. (\ref{exs22}%
), and integral (\ref{3.3}) can be represented as:%
\begin{eqnarray}
N^{\mathrm{cr}} &\approx &\frac{V_{\bot }TJ_{(d)}}{(2\pi )^{d-1}}%
\int_{\alpha \pi _{\bot }<K_{\bot }}\left( I_{p_{\bot }}^{+}+I_{p_{\bot
}}^{-}\right) d\mathbf{p}_{\bot }\ ,  \notag \\
I_{p_{\bot }}^{+} &=&\int_{0}^{eE_{0}\sigma /\sqrt{3}-K/\sigma }e^{-\pi \tau
}d\pi _{0}^{\prime }\ ,  \notag \\
I_{p_{\bot }}^{-} &=&\int_{-\left[ eE_{0}\sigma \left( 1-1/\sqrt{3}\right)
-K/\sigma \right] }^{0}e^{-\pi \tau }d\pi _{0}^{\prime }\ ,  \label{exs33}
\end{eqnarray}%
where $\tau $\ is given by Eq. (\ref{exs22}). To calculate $I_{p_{\bot
}}^{+} $\ and $I_{p_{\bot }}^{-}$, it is convenient to use the
representation $\tau =\lambda \left( q+1\right) $ and to pass from the
integration over $\pi _{0}^{\prime }$ to the integration over the parameter $%
q$ (the transition to such a variable provides exponential decrease of the
integrand with increasing $q$, and the expansion of the pre-exponential
factor in powers of $q$ has a form of an asymptotic series).

Finding $\pi _{0}^{\prime }$ as a function of $q$, one has to take into
account that in the region $D$ the following expansions are valid up to
linear terms in reciprocal powers of large parameters:%
\begin{eqnarray*}
&&\left\vert p^{\mathrm{L}}\right\vert \approx \frac{1}{\epsilon _{1}}-\frac{%
\epsilon _{1}\pi _{\perp }^{2}}{2}+O(\epsilon _{1}),\ \ \epsilon _{1}=\left[
\pi _{0}^{^{\prime }}+eE_{0}\sigma \left( 1-1/\sqrt{3}\right) \right] ^{-1}\
, \\
&&\left\vert p^{\mathrm{R}}\right\vert \approx \frac{1}{\epsilon _{2}}-\frac{%
\epsilon _{2}\pi _{\perp }^{2}}{2}+O(\epsilon _{2}),\ \ \epsilon _{2}=\left[
eE_{0}\sigma /\sqrt{3}-\pi _{0}^{^{\prime }}\right] ^{-1}\ , \\
&&\sqrt{2\delta U^{2}+2\left\vert p^{\mathrm{R}}\right\vert ^{2}-\left\vert
p^{\mathrm{L}}\right\vert ^{2}}\approx \frac{1}{\epsilon _{3}}-\frac{%
\epsilon _{3}\pi _{\perp }^{2}}{2}+O(\epsilon _{3}),\ \ \epsilon _{0}=\left[
eE_{0}\sigma \left( 1+1/\sqrt{3}\right) -\pi _{0}^{^{\prime }}\right] ^{-1}\
.
\end{eqnarray*}%
Then we obtain the following approximation for $\tau :$%
\begin{equation}
\tau \approx \frac{2\sqrt{3}\lambda }{2\sqrt{3}+9\left( r-\sqrt{3}\right)
r^{2}},\ \ r=\frac{\pi _{0}^{^{\prime }}}{eE_{0}\sigma }\ .  \label{3.20}
\end{equation}

Besides, we have to find $\pi _{0}^{\prime }$ as a function of $q$ using Eq.
(\ref{3.20}). Such a function can be found from the cubic equation%
\begin{equation}
r^{3}-\sqrt{3}r^{2}+\frac{2q}{3\sqrt{3}\left( q+1\right) }=0\ .  \label{3.31}
\end{equation}%
Note that when $\pi _{0}^{^{\prime }}\rightarrow eE_{0}\sigma /\sqrt{3}%
-K/\sigma $\ and $\pi _{0}^{^{\prime }}\rightarrow -\left[ eE_{0}\sigma
\left( 1-1/\sqrt{3}\right) -K/\sigma \right] $, the parameter $\tau $\
reaches the limiting values $\tau _{\max }^{\pm }=\lambda \left( q_{\max
}^{\pm }+1\right) $, respectively. However, since contributions of the
factor $\exp \left( -\pi \tau \right) $\ to integrals (\ref{exs33}) outside
of range $D$ are exponentially small, one can extend limits of the
integration over $q$\ to $\pm \infty $. Equation (\ref{3.31}) has three real
roots:%
\begin{align}
& r_{1}=\frac{2}{\sqrt{3}}\cos \frac{\alpha \left( q\right) }{3}+1/\sqrt{3}%
,\ \alpha \left( q\right) =\arccos \left[ \left( q+1\right) ^{-1}\right] \ ,
\notag \\
& r_{2}=-\frac{2}{\sqrt{3}}\cos \left[ \frac{\alpha \left( q\right) }{3}+%
\frac{\pi }{3}\right] +1/\sqrt{3}\ ,  \notag \\
& r_{3}=-\frac{2}{\sqrt{3}}\cos \left[ \frac{\alpha \left( q\right) }{3}-%
\frac{\pi }{3}\right] +1/\sqrt{3}\ ;  \label{3.33}
\end{align}%
see, e.g., \cite{Korn}.

Since $0<q<+\infty $, the inequality $0\leq \alpha \left( q\right) \leq \pi
/2$ holds true, which implies: 
\begin{equation}
0\leq \frac{\alpha \left( q\right) }{3}\leq \frac{\pi }{6},\ \ \frac{\pi }{3}%
\leq \left[ \frac{\alpha \left( q\right) }{3}+\frac{\pi }{3}\right] \leq 
\frac{\pi }{2},\ \ -\frac{\pi }{3}\leq \left[ \frac{\alpha \left( q\right) }{%
3}-\frac{\pi }{3}\right] \leq -\frac{\pi }{6}\ .  \label{3.36}
\end{equation}%
Then 
\begin{equation}
1\leq r_{1}\leq \sqrt{3},\ \ 0\leq r_{2}\leq \frac{1}{\sqrt{3}},\ \ \left( 
\frac{1}{\sqrt{3}}-1\right) \leq r_{3}\leq 0  \label{3.38}
\end{equation}%
such that roots $r_{2}$\ and $r_{3}$\ represent the quantity $\pi
_{0}^{^{\prime }}$ in the subranges\ $\pi _{0}^{^{\prime }}\in \left(
-eE_{0}\sigma \left( 1-1/\sqrt{3}\right) ,0\right) \ $and $\pi
_{0}^{^{\prime }}\in \left( 0,eE_{0}\sigma /\sqrt{3}\right) $, respectively.

Thus, the integrals $I_{p_{\bot }}^{+}$ and $I_{p_{\bot }}^{-}$ take the
forms:%
\begin{equation}
I_{p_{\perp }}^{\pm }=\pm \frac{2\delta U}{3\sqrt{3}}\int_{0}^{+\infty }dq%
\frac{\left( q+1\right) ^{-2}}{\sqrt{1-\left( q+1\right) ^{-2}}}\sin \left[ 
\frac{\alpha \left( q\right) }{3}\pm \frac{\pi }{3}\right] \exp \left[ -\pi
\lambda \left( q+1\right) \right] \ ,  \label{3.39}
\end{equation}%
and their sum can be represented as:%
\begin{equation}
I_{p_{\perp }}^{-}+I_{p_{\perp }}^{+}=\frac{2\delta U}{3}\int_{0}^{+\infty
}dq\frac{\left( 1+q\right) ^{-2}}{\sqrt{1-\left( 1+q\right) ^{-2}}}\cos 
\frac{\alpha \left( q\right) }{3}\exp \left[ -\pi \lambda \left( q+1\right) %
\right] \ .  \label{3.40}
\end{equation}

Substituting Eq. (\ref{3.40}) into Eq. (\ref{exs33}) and integrating over $%
dp_{\perp }^{\left( d-2\right) }$, we obtain:%
\begin{align}
& N^{\mathrm{cr}}\approx V_{\bot }T\rho ^{\mathrm{D}},\;\rho ^{\mathrm{D}%
}=\beta \frac{\delta U}{eE_{\max }}w,\ \ \beta =\frac{J_{\left( d\right) }%
\left[ eE_{\max }\right] ^{d/2}}{\left( 2\pi \right) ^{d-1}}\exp \left[ -%
\frac{\pi m^{2}}{eE_{\max }}\right] \ ,  \notag \\
& w\left( d\right) =\frac{2}{3}\int_{0}^{+\infty }dq\left( q^{2}+2q\right)
^{-1/2}\left( q+1\right) ^{-d/2}\cos \frac{\alpha \left( q\right) }{3}\exp %
\left[ -\frac{\pi m^{2}}{eE_{\max }}q\right] \ .  \label{3.41}
\end{align}%
The corresponding probability $P_{\mathrm{v}}$ of the vacuum to remain a
vacuum reads:%
\begin{equation}
P_{\mathrm{v}}=\exp \left[ -\mu N^{\mathrm{cr}}\right] ,\ \ \mu
=\sum_{l=0}^{\infty }\left( l+1\right) ^{-d/2}\exp \left( -l\frac{\pi m^{2}}{%
eE_{\max }}\right) \ .  \label{3.42}
\end{equation}

In general, the current density $J_{x}^{\mathrm{cr}}$ of created particles
is given by Eq. (\ref{d19b}), while the charge polarization due to the
separation of created positrons and electrons is presented by the charge
density $\rho ^{\mathrm{cr}}\left( x\right) $ given by Eq. (\ref{d22}). The
nonzero components of energy-momentum tensor of the created particles are
given by Eq. (\ref{d23}). In the subrange $D\subset \Omega _{3}$ the
velocities $v^{\mathrm{L}}$ and $v^{\mathrm{R}}$ tend to the speed of the
light $c=1$ such that $\left\vert p^{\mathrm{L/R}}\right\vert \approx
\left\vert \pi _{0}^{\prime }\left( \mathrm{L/R}\right) \right\vert \approx
\left\vert U_{\mathrm{L/R}}^{\prime }\right\vert $, with $U_{\mathrm{L}%
}^{\prime }=-eE_{0}\sigma \left( 1-1/\sqrt{3}\right) $ and $U_{\mathrm{R}%
}^{\prime }=eE_{0}\sigma /\sqrt{3}$ according to inequality (\ref{exs24}).
Using representation (\ref{3.41}), we find:%
\begin{eqnarray}
&&J_{x}^{\mathrm{cr}}\approx e\rho ^{\mathrm{D}},\;J_{0}^{\mathrm{cr}}\left(
x\right) \approx \left\{ 
\begin{array}{c}
-J_{x}^{\mathrm{cr}},\ \ x\in S_{\mathrm{L}} \\ 
J_{x}^{\mathrm{cr}},\ x\in S_{\mathrm{R}}%
\end{array}%
\right. \ ;  \notag \\
&&T_{\mathrm{cr}}^{00}(x)\approx T_{\mathrm{cr}}^{11}(x)\approx \left\{ 
\begin{array}{c}
\rho ^{\mathrm{D}}\left\vert U_{\mathrm{L}}^{\prime }\right\vert ,\ x\in S_{%
\mathrm{L}} \\ 
\rho ^{\mathrm{D}}\left\vert U_{\mathrm{R}}^{\prime }\right\vert ,\mathrm{\ }%
x\in S_{\mathrm{R}}%
\end{array}%
\right. \ ,  \notag \\
&&T_{\mathrm{cr}}^{kk}(x)\approx \left\{ 
\begin{array}{c}
\tilde{\rho}^{\mathrm{D}}\left\vert U_{\mathrm{L}}^{\prime }\right\vert
^{-1},\mathrm{\ }x\in S_{\mathrm{L}} \\ 
\tilde{\rho}^{\mathrm{D}}\left\vert U_{\mathrm{R}}^{\prime }\right\vert
^{-1},\mathrm{\ }x\in S_{\mathrm{R}}%
\end{array}%
\right. ,\ \ k\neq 1\ ,  \notag \\
&&T_{\mathrm{cr}}^{10}(x)\approx \left\{ 
\begin{array}{c}
-\rho ^{\mathrm{D}}\left\vert U_{\mathrm{L}}^{\prime }\right\vert ,\ x\in S_{%
\mathrm{L}} \\ 
\rho ^{\mathrm{D}}\left\vert U_{\mathrm{R}}^{\prime }\right\vert ,\mathrm{\ }%
x\in S_{\mathrm{R}}%
\end{array}%
\right. \ ,\ \tilde{\rho}^{\mathrm{D}}=\beta \frac{\delta U\mathbb{\sigma }}{%
2\pi }w\left( d+2\right) .  \label{gav1}
\end{eqnarray}%
It can be seen that the transversal components of the pressure of the
created particles are much less than the longitudinal pressure in the areas $%
S_{\mathrm{L}}$ and $S_{\mathrm{R}}$, respectively, $T_{\mathrm{cr}%
}^{kk}(x)\ll T_{\mathrm{cr}}^{11}(x)$. It is in accordance with the relation 
$T_{\mathrm{cr}}^{00}(x)\approx T_{\mathrm{cr}}^{11}(x)$.

Unlike the case with the Sauter field, the energy density and the
longitudinal component of the pressure in the areas $S_{\mathrm{L}}$\ and $%
S_{\mathrm{R}}$, given by Eq. (\ref{gav1}), are not equal, $T_{\mathrm{cr}%
}^{00}(x_{\mathrm{L}})=2T_{\mathrm{cr}}^{00}(x_{\mathrm{R}})$\ and $T_{%
\mathrm{cr}}^{11}(x_{\mathrm{L}})=2T_{\mathrm{cr}}^{11}(x_{\mathrm{R}})$,
that is, the energy density and longitudinal pressure of created electrons
is two times more than the energy density and longitudinal pressure of
created positrons due to the asymmetric form of the field (\ref{1.2}). The
direction of this field is chosen along the $x$-axis. Choosing the opposite
direction of the field, one finds that the energy density and longitudinal
pressure of created positrons is two times more than the energy density and
the longitudinal pressure of created electrons.

One can show that representations (\ref{3.41}) - (\ref{gav1}) can be
obtained in the framework of a new kind of a locally constant field
approximation (LCFA) constructed in Ref. \cite{GavGitShi19b}; see Appendix %
\ref{Ap2}. This can be considered as an additional argument in favor of the
validity of such an approximation.

Besides, it turns out that LCFA allows one to see relation to the
Heisenberg-Euler effective action method. Note that in the framework of the
LCFA the probability $P_{v}$, given by Eq. (\ref{3.42}),\ can be represented
via the imaginary part of a one-loop effective action $S$\ by the seminal
Schwinger formula \cite{Schw51},%
\begin{equation}
P_{\mathrm{v}}=\exp \left( -2\mathrm{Im}S\right) .  \label{np1a}
\end{equation}%
Thus, all the vacuum mean values, obtained in the framework of the LCFA, can
be associated with the effective action approach. If the total number of
created particles is small, $N^{\mathrm{cr}}\ll 1$, then $1-P_{v}\approx N^{%
\mathrm{cr}}$\emph{. }Therefore, knowledge of the probability $P_{v}$\
allows one to estimate the total number of created particles $N^{\mathrm{cr}%
} $. It is, in this case, the effective action approach to calculating $%
P_{v} $\ turns out to be useful. We note that this\ approach is a base of a
number of approximation methods; see, e.g., Ref. \cite{Adv-QED22} for a
review. In this relation, it should be noted that the probability $P_{v}$\
by itself is not very useful in the case of strong fields when $P_{v}\ll 1$.
In the latter case it is necessary to directly calculate the vacuum mean
values of physical quantities, using either exact solutions, as it is done
above, or the new kind of the LCFA \cite{GavGitShi19b}.

\section{The Klein step\label{S6}}

The Klein paradox is known from the work by Klein \cite{Klein27} who
considered, in the framework of the one-particle relativistic theory,
reflection and transmission probabilities of charged relativistic particles
incident on a sufficiently high rectangular potential step (the Klein step)
of the form%
\begin{equation}
qA_{0}\left( x\right) =\left\{ 
\begin{array}{l}
U_{\mathrm{L}},\ x<0 \\ 
U_{\mathrm{R}},\ x>0%
\end{array}%
\right. \ ,  \label{exs41}
\end{equation}%
where $U_{\mathrm{R}}$ and $U_{\mathrm{L}}$ are some constants. The field (%
\ref{exs41}) represents a kind of $x$-step. According to calculations of
Klein and other authors, for certain energies and sufficient high magnitude $%
U=U_{\mathrm{R}}-U_{\mathrm{L}}$ of the Klein step,\ it seems that there are
more reflected fermions than incident. This fact many articles and books
where treated as a paradox (the Klein paradox); see Refs. \cite%
{DomCal99,HansRavn81}) for historical review.\emph{\ }This paradox and other
misunderstandings in considering quantum effects in fields of strong $x$%
-steps can be consistently solved as many particle effects in the $QED$\
with an unstable vacuum; see \cite{GavGi16}. Obviously that the Klein step
is a limiting case of a very sharp peak field. It is important to have in
hands examples of potentials representing very sharp peak field that can be
considered as their regularizations. Such an example given by Sauter
potential (\ref{1.4}) was presented in Ref. \cite{GavGi16}; see Refs. \cite%
{GavGitSh17,AdoGavGit20} for regularizations by piecewise forms of analytic
functions. The $x$-step (\ref{1.2}) under consideration represents a new
example of such regularizations given by an analytic function.

Let us study characteristics of the vacuum instability caused by the field (%
\ref{1.2}) with $\sigma $ sufficiently small, $\sigma \rightarrow 0$\emph{.}
If $U_{\mathrm{R}}=0$ and $U_{\mathrm{L}}=-\delta U=\mathbb{-}eE_{0}\sigma $
are given constant and 
\begin{equation}
\delta U\sigma \ll 1\   \label{exs42}
\end{equation}%
the field imitates sufficiently well the asymmetric Klein step (\ref{exs41})
and coincides with the latter as $\sigma \rightarrow 0$.

In the ranges $\Omega _{1}$ and $\Omega _{5}$ the energy $\left\vert
p_{0}\right\vert $ is not restricted from the above, that is why, in what
follows, we consider only the subranges, where$\max \left\{ \sigma
\left\vert p^{\mathrm{L}}\right\vert ,\sigma \left\vert p^{\mathrm{R}%
}\right\vert \right\} \ll 1$. In the leading-term approximation in\emph{\ }$%
\sigma $ it follows from Eqs.~(\ref{2.20}) that%
\begin{equation}
\left\vert g\left( _{+}\left\vert ^{-}\right. \right) \right\vert
^{-2}\approx \frac{4k}{\left( 1-k\right) ^{2}},\ \ k=\frac{\left\vert p^{%
\mathrm{R}}\right\vert }{\left\vert p^{\mathrm{L}}\right\vert }\frac{\pi
_{0}\left( \mathrm{L}\right) +\pi _{\bot }}{\pi _{0}\left( \mathrm{R}\right)
+\pi _{\bot }}\ ,  \label{exs45}
\end{equation}%
\ where $k$ is called the kinematic factor.

Note that $k$ are positive and do not achieve the unit values, $k\neq 1$ in
the ranges $\Omega _{1}$ and $\Omega _{5}$. In these ranges, the
coefficients $g$ satisfy the same relations,%
\begin{equation}
\left\vert g\left( _{+}\left\vert ^{+}\right. \right) \right\vert
^{2}=\left\vert g\left( _{+}\left\vert ^{-}\right. \right) \right\vert
^{2}+1\ .  \label{exs47}
\end{equation}%
Therefore, reflection and transmission probabilities derived from Eqs.~(\ref%
{exs45}) have the same forms%
\begin{eqnarray}
\left\vert T_{\zeta ,n}\right\vert ^{2} &=&\left\vert g\left( _{+}\left\vert
^{+}\right. \right) \right\vert ^{-2}=\frac{4k}{\left( 1+k\right) ^{2}}, 
\notag \\
\left\vert R_{\zeta ,n}\right\vert ^{2} &=&\left\vert g\left( _{+}\left\vert
^{-}\right. \right) \right\vert ^{2}\left\vert g\left( _{+}\left\vert
^{+}\right. \right) \right\vert ^{-2}=\frac{\left( 1-k\right) ^{2}}{\left(
1+k\right) ^{2}}\ .  \label{exs49}
\end{eqnarray}

To compare our exact results with results of the nonrelativistic
consideration obtained in any textbook for one dimensional quantum motion,
we set $p_{\bot }=0,$\ then $\pi _{\bot }=m,$ $\pi _{0}\left( \mathrm{L}%
\right) =m+E$, and $\pi _{0}\left( \mathrm{R}\right) =m+E-\delta U$. In the
nonrelativistic limit, when $\delta U,E\ll m$, we obtain%
\begin{equation*}
k=k^{\mathrm{NR}}=\sqrt{\frac{E-\delta U}{E}}\ ,
\end{equation*}%
which can be identified with the nonrelativistic results, e.g., see \cite%
{LanLiQM}.

Let us consider the range $\Omega _{3}$. Here the quantum numbers $\mathbf{p}%
_{\bot }$ are restricted by the inequality$\ 2\pi _{\bot }\leq \delta U$ and
for any of such $\pi _{\bot }$\ the quantum numbers $p_{0}$ obey the strong
inequality (\ref{3.0}). In this range the quantity $\left\vert g\left(
_{+}\left\vert ^{-}\right. \right) \right\vert ^{-2}$ represents the
differential mean numbers of electron-positron pairs created from the
vacuum, $N_{n}^{\mathrm{cr}}=\left\vert g\left( _{+}\left\vert ^{-}\right.
\right) \right\vert ^{-2}$. In this range for any given $\pi _{\bot }$ the
absolute values of $\left\vert p^{\mathrm{R}}\right\vert $ and $\left\vert
p^{\mathrm{L}}\right\vert $ are restricted from the above, see (\ref{d8}).
Therefore, condition (\ref{exs42}) implies Eq. (\ref{exs43}). It follows
from Eq.~(\ref{2.20}) that in the leading approximation the following
equation holds true%
\begin{equation}
\left\vert g\left( _{+}\left\vert ^{-}\right. \right) \right\vert
^{-2}\approx \frac{4\left\vert p^{\mathrm{L}}\right\vert \left\vert p^{%
\mathrm{R}}\right\vert }{\delta U^{2}-\left( \left\vert p^{\mathrm{L}%
}\right\vert -\left\vert p^{\mathrm{R}}\right\vert \right) ^{2}}=\frac{%
4\left\vert k\right\vert }{\left( 1+\left\vert k\right\vert \right) ^{2}}.
\label{exs50}
\end{equation}

Note that expression (\ref{exs50}) differs from expression (\ref{exs45})
only by the sign of the kinematic factor $k$. This factor is positive in the
ranges $\Omega _{1}$ and $\Omega _{5}$, and it is negative in the range $%
\Omega _{3}$. In the range $\Omega _{3}$, the difference $\left\vert p^{%
\mathrm{L}}\right\vert -\left\vert p^{\mathrm{R}}\right\vert $ may be zero
at $p_{0}=U_{\mathrm{L}}/2$, which corresponds to $k=-\left( \delta U+2\pi
_{\bot }\right) /\left( \delta U-2\pi _{\bot }\right) $. Namely in this case
the quantity $\left\vert g\left( _{+}\left\vert ^{-}\right. \right)
\right\vert ^{-2}$ has a maximum at a given $\pi _{\bot }$,%
\begin{equation}
\max \left\vert g\left( _{+}\left\vert ^{-}\right. \right) \right\vert
^{-2}=1-\left( 2\pi _{\bot }/\delta U\right) ^{2}.  \label{exs50b}
\end{equation}

Note that expressions (\ref{exs45}), (\ref{exs49}) and (\ref{exs50})
coincide up to the redesigning of the constants $U_{\mathrm{L}/\mathrm{R}}$
with expressions, corresponding to other regularizations of the Klein step;
see Ref. \cite{GavGi16,GavGitSh17,AdoGavGit20}. We see that the Klein step
is (in a sense) a limiting case of various sharp peak fields under condition
that the magnitude $\delta U$ is a given finite constant and the peak of the
fields are sufficiently sharp.

\section{Concluding remarks\label{S7}}

A new exactly solvable case in strong-field $QED$ with $x$-step is
presented. This step can be seen as a certain analytic "deformation" of the
Sauter field. In contrast to the Sauter field the potential field under
consideration is asymmetric with respect of the axis $x$ reflection. Bearing
in mind numerous examples of using the results of the exactly solvable
problem with the Sauter field in physical applications related to the
problem of vacuum instability, we believe that the new exact solvable case
will also be useful in such applications. It can be treated as a new
regularization of the Klein step. Exact solutions of the Dirac equation used
in the above nonperturbative calculations, are presented in the form of
stationary plane waves with special left and right asymptotics and
identified as components of initial and final wave packets of particles and
antiparticles in the framework of the strong-field $QED$ \cite{GavGi16}. We
show that in spite of the fact that the symmetry with respect to positive
and negative bands of energies is broken, distribution of created pairs and
other physical quantities can be presented by elementary functions. We
consider the processes of transmission and reflection in the ranges of the
stable vacuum and study physical quantities specifying the vacuum
instability as well. We find the differential mean numbers of
electron-positron pairs created from the vacuum, the components of current
density and energy-momentum tensor of the created electrons and positrons
leaving the area of the strong field under consideration. We study the
particular case of the particle creation due to a weakly inhomogeneous
electric field and obtain explicitly the total number, the current density
and energy-momentum tensor of created particles. Unlike the symmetric case
of the Sauter field the asymmetric form of the field under consideration
causes the energy density and longitudinal pressure of created electrons to
be not equal to the energy density and longitudinal pressure of created
positrons.

\begin{acknowledgments}
 Gitman is grateful to CNPq for continued support.
\end{acknowledgments}

\appendix

\section{Basic elements of a nonperturbative approach to $QED$ with $x$%
-steps \label{Ap}}

In this appendix, we briefly present some basic constructions of
quantization in terms of particles for QED with $x$-steps; see sections
IV-VII in Ref. \cite{GavGi16} for details.

The time-independent inner product for any pair of solutions of the Dirac
equation, $\psi _{n}\left( X\right) $ and $\psi _{n^{\prime }}^{\prime
}\left( X\right) $, is defined on the $t=$\textrm{const} hyperplane as
follows: 
\begin{equation}
\left( \psi _{n},\psi _{n^{\prime }}^{\prime }\right) =\int_{V_{\bot }}d%
\mathbf{r}_{\bot }\int\limits_{-K^{\left( \mathrm{L}\right) }}^{K^{\left( 
\mathrm{R}\right) }}\psi _{n}^{\dag }\left( X\right) \psi _{n^{\prime
}}^{\prime }\left( X\right) dx,\ \   \label{t4}
\end{equation}%
where the integral over the spatial volume $V_{\bot }${\large \ }is
completed by an integral\ over the interval $\left[ K^{\left( \mathrm{L}%
\right) },K^{\left( \mathrm{R}\right) }\right] $ in the $x$ direction. The
parameters $K^{\left( \mathrm{L}/\mathrm{R}\right) }$ are assumed
sufficiently large in final expressions. Assuming that the principal value
of integral (\ref{t4}) is determined by integrals over the areas where the
field $E(x)$ is negligible small it is possible to evaluate this value using
only the asymptotic behavior (\ref{2.16}) of functions in the regions where
particles are free. The field $E(x)$ in the area where it is strong enough
affects only coefficients $g$ entering into the mutual decompositions of the
solutions given by Eq. (\ref{2.25}). One can see that the norms of the plane
waves $\ _{\zeta }\psi _{n}\left( X\right) $ and $\ ^{\zeta }\psi _{n}\left(
X\right) $ with respect to the inner product (\ref{t4}) are proportional to
the macroscopically large parameters{\large \ }$\tau ^{\left( \mathrm{L}%
\right) }=K^{\left( \mathrm{L}\right) }/v^{\mathrm{L}}${\large \ }and{\large %
\ }$\tau ^{\left( \mathrm{R}\right) }=K^{\left( \mathrm{R}\right) }/v^{%
\mathrm{R}}${\large , }where $v^{\mathrm{L}}=\left\vert p^{\mathrm{L}}/\pi
_{0}\left( \mathrm{L}\right) \right\vert >0$ and $v^{\mathrm{R}}=\left\vert
p^{\mathrm{R}}/\pi _{0}\left( \mathrm{R}\right) \right\vert >0$ are absolute
values of the longitudinal velocities of particles in the regions where
particles are free; see Sect. IIIC.2 and Appendix B in Ref. \cite{GavGi16}
for details.

It was shown (see Appendix B in Ref. \cite{GavGi16}) that\ the following
couples of plane waves are orthogonal with \ respect to the inner product (%
\ref{t4})%
\begin{equation}
\left( _{\zeta }\psi _{n},^{-\zeta }\psi _{n}\right) =0,\ \ n\in \Omega
_{1}\cup \Omega _{5}\ ;\ \ \left( _{\zeta }\psi _{n},^{\zeta }\psi
_{n}\right) =0,\ \ n\in \Omega _{3}\ ,  \label{i7}
\end{equation}%
if the parameters of the volume regularization $\tau ^{\left( \mathrm{L/R}%
\right) }$\ satisfy the condition%
\begin{equation}
\tau ^{\left( \mathrm{L}\right) }-\tau ^{\left( \mathrm{R}\right) }=O\left(
1\right) ,  \label{i8}
\end{equation}%
where $O\left( 1\right) $ denotes terms that are negligibly small in
comparison with the macroscopic quantities $\tau ^{\left( \mathrm{L/R}%
\right) }${\large . }One can see that $\tau ^{\left( \mathrm{R}\right) }$
and $\tau ^{\left( \mathrm{L}\right) }$ are macroscopic times of motion of
particles and antiparticles in the areas $S_{\mathrm{R}}$ and $S_{\mathrm{L}%
} $, respectively and they are equal,{\large \ }%
\begin{equation}
\tau ^{\left( \mathrm{L}\right) }=\tau ^{\left( \mathrm{R}\right) }=\tau .
\label{m4}
\end{equation}%
It allows one to introduce an unique time of motion $\tau $\ for all the
particles in the system under consideration. This time can be interpreted as
a system monitoring time during its evolution and as such is fixed as $\tau
=T$ in the framework of \ the renormalization procedure; see Ref. \cite%
{GavGi20} for details.

The renormalization and volume regularization procedures are{\large \ }%
associated with the introduction of a modified inner product and a parameter 
$\tau $ of the regularization. Based on physical considerations, we fix this
parameter. It turns out that in the Klein range this parameter can be
interpreted as the time of the observation of the pair production process.

Under condition (\ref{i8}) the following orthonormality relations on the $t=$%
\textrm{const} hyperplane are 
\begin{eqnarray}
&&\left( \ _{\zeta }\psi _{n},\ _{\zeta }\psi _{n^{\prime }}\right) =\left(
\ ^{\zeta }\psi _{n},\ ^{\zeta }\psi _{n^{\prime }}\right) =\delta
_{n,n^{\prime }}\mathcal{M}_{n}\ ,\ \ n\in \Omega _{1}\cup \Omega _{3}\cup
\Omega _{5}\ ;  \notag \\
&&\left( \psi _{n},\psi _{n^{\prime }}\right) =\delta _{n,n^{\prime }}%
\mathcal{M}_{n},\;n\in \Omega _{2}\cup \Omega _{4}\ ,  \notag \\
&&\mathcal{M}_{n}=2\frac{\tau }{T}\left\vert g\left( _{+}\left\vert
^{+}\right. \right) \right\vert ^{2},\ \ n\in \Omega _{1}\cup \Omega _{5}, 
\notag \\
&&\mathcal{M}_{n}=2\frac{\tau }{T}\left\vert g\left( _{+}\left\vert
^{-}\right. \right) \right\vert ^{2},\ \ n\in \Omega _{3},  \notag \\
&&\mathcal{M}_{n}=2\frac{\tau }{T},\ \ n\in \Omega _{2},\mathrm{\;}\mathcal{M%
}_{n}=2\frac{\tau }{T},\ \ n\in \Omega _{4}.  \label{i12}
\end{eqnarray}%
All the wave functions having different quantum numbers $n$ are orthogonal,
and 
\begin{eqnarray}
&&\left( \ _{\zeta }\psi _{n},\ ^{-\zeta }\psi _{n}\right) =0,\ \ n\in
\Omega _{1}\cup \Omega _{5}\ ,\ _{\ \zeta }\psi _{n}\ \mathrm{and}\ \
^{-\zeta }\psi _{n}\ \mathrm{independent,}  \notag \\
&&\left( \ _{\zeta }\psi _{n},\ ^{\zeta }\psi _{n}\right) =0,\ \ n\in \Omega
_{3}\ ,\ _{\zeta }\psi _{n}\ \mathrm{and}\ \ ^{\zeta }\psi _{n}\ \mathrm{%
independent.}  \label{i9}
\end{eqnarray}%
We denote\emph{\ }the corresponding quantum numbers by $n_{k}$,\ so that $%
n_{k}\in \Omega _{k}$. Then we identify components of the initial and final
wave packets of particles and antiparticles in Eq. (\ref{in-out}).

We decompose the Heisenberg operator $\hat{\Psi}\left( X\right) $ in two
sets of solutions $\left\{ \ _{\zeta }\psi _{n}\left( X\right) \right\} $
and $\left\{ \ ^{\zeta }\psi _{n}\left( X\right) \right\} $ of the Dirac
equation (\ref{2.6}) complete on the $t=\mathrm{const}$ hyperplane.
Operator-valued coefficients in such decompositions do not depend on
coordinates. Our division of the quantum numbers $n$ in five ranges $\Omega
_{k}$, implies the representation for $\hat{\Psi}\left( X\right) $ as a sum
of five operators $\hat{\Psi}_{k}\left( X\right) $, $k=1,2,3,4,5$,%
\begin{equation}
\hat{\Psi}\left( X\right) =\sum_{k=1}^{5}\hat{\Psi}_{k}\left( X\right) .
\label{2.20b}
\end{equation}

For each of three operators $\hat{\Psi}_{k}\left( X\right) $,$\ k=1,3,5,$
there exist two possible decompositions (\ref{in-out}) according to the
existence of two different complete sets of solutions with the same quantum
numbers $n$ in the ranges $\Omega _{1}$, $\Omega _{3}$, and $\Omega _{5}$,%
\begin{eqnarray}
\hat{\Psi}_{1}\left( X\right) &=&\sum_{n_{1}}\mathcal{M}_{n_{1}}^{-1/2}\left[
\ _{+}a_{n_{1}}(\mathrm{in})\ _{+}\psi _{n_{1}}\left( X\right) +\
^{-}a_{n_{1}}(\mathrm{in})\ ^{-}\psi _{n_{1}}\left( X\right) \right]  \notag
\\
&=&\sum_{n_{1}}\mathcal{M}_{n_{1}}^{-1/2}\left[ \ ^{+}a_{n_{1}}(\mathrm{out%
})\ ^{+}\psi _{n_{1}}\left( X\right) +\ _{-}a_{n_{1}}(\mathrm{out})\
_{-}\psi _{n_{1}}\left( X\right) \right] ,  \notag \\
\hat{\Psi}_{3}\left( X\right) &=&\sum_{n_{3}}\mathcal{M}_{n_{3}}^{-1/2}\left[
\ ^{-}a_{n_{3}}(\mathrm{in})\ ^{-}\psi _{n_{3}}\left( X\right) +\
_{-}b_{n_{3}}^{\dagger }(\mathrm{in})\ _{-}\psi _{n_{3}}\left( X\right) %
\right]  \notag \\
&=&\sum_{n_{3}}\mathcal{M}_{n_{3}}^{-1/2}\left[ \ ^{+}a_{n_{3}}(\mathrm{out%
})\ ^{+}\psi _{n_{3}}\left( X\right) +\ _{+}b_{n_{3}}^{\dagger }(\mathrm{out}%
)\ _{+}\psi _{n_{3}}\left( X\right) \right] ,  \notag \\
\hat{\Psi}_{5}\left( X\right) &=&\sum_{n_{5}}\mathcal{M}_{n_{5}}^{-1/2}\left[
\ ^{+}b_{n_{5}}^{\dag }(\mathrm{in})\ ^{+}\psi _{n_{5}}\left( X\right) +\
_{-}b_{n_{5}}^{\dag }(\mathrm{in})\ _{-}\psi _{n_{5}}\left( X\right) \right]
\notag \\
&=&\sum_{n_{5}}\mathcal{M}_{n_{5}}^{-1/2}\left[ \ _{+}b_{n_{5}}^{\dag }(%
\mathrm{out})\ _{+}\psi _{n_{5}}\left( X\right) +\ ^{-}b_{n_{5}}^{\dag }(%
\mathrm{out})\ ^{-}\psi _{n_{5}}\left( X\right) \right] .  \label{2.23}
\end{eqnarray}

There may exist only one complete set of solutions with the same quantum
numbers $n_{2}$ and $n_{4}$. Therefore, we have only one possible
decomposition for each of the\ two operators $\hat{\Psi}_{i}\left( X\right)
, $ $i=2,4$,%
\begin{equation}
\hat{\Psi}_{2}\left( X\right) =\sum_{n_{2}}\mathcal{M}%
_{n_{2}}^{-1/2}a_{n_{2}}\psi _{n_{2}}\left( X\right) ,\ \ \hat{\Psi}%
_{4}\left( X\right) =\sum_{n_{4}}\mathcal{M}_{n_{4}}^{-1/2}b_{n_{4}}^{%
\dagger }\psi _{n_{4}}\left( X\right) .  \label{2.21}
\end{equation}

We interpret all $a$ and $b\ $as annihilation and all $a^{\dag }$ and $%
b^{\dag }$ as creation operators. All $a$ and $a^{\dag }$ are interpreted\
as describing electrons and all $b$ and $b^{\dag }$ as describing positrons.
All the operators labeled by the argument \textrm{in} are interpreted\ as 
\textrm{in}-operators, whereas all the operators labeled by the argument 
\textrm{out} as \textrm{out}-operators. This identification is confirmed by
a detailed mathematical and physical analysis of solutions of the Dirac
equation with subsequent QFT analysis of correctness of such an
identification in Ref. \cite{GavGi16}.

Taking into account the orthogonality and orthonormalization relations, we
find that the standard anticommutation relations for the Heisenberg operator
(\ref{2.20b}) yield the standard anticommutation rules for the introduced
creation and annihilation \textrm{in}- or \textrm{out-}operators.

We define two vacuum vectors $\left\vert 0,\mathrm{in}\right\rangle $ and $%
\left\vert 0,\mathrm{out}\right\rangle $, one of which is the\ zero-vector
for all \textrm{in}-annihilation operators and the other is zero-vector for
all $\mathrm{out}$-annihilation operators. Besides, both vacua are
zero-vectors for the annihilation operators $a_{n_{2}}$ and $b_{n_{4}}$. One
can verify that the introduced vacua have minimum (zero by definition)
kinetic energy and zero electric charge and all the excitations above the
vacuum have positive energies. Then we postulate that the state space of the
system under consideration is the Fock space constructed, say, with the help
of the vacuum $\left\vert 0,\mathrm{in}\right\rangle $ and the corresponding
creation operators. This Fock space is unitarily equivalent to the\ other
Fock space constructed with the help of the vacuum $\left\vert 0,\mathrm{out}%
\right\rangle $ and the corresponding creation operators if the total number
of particles created by the external field is finite.

Because any annihilation operators with quantum numbers $n_{k}$
corresponding to different $k$ anticommute between themselves, we can
represent the introduced vacua $\left\vert 0,\mathrm{in}\right\rangle $ and $%
\left\vert 0,\mathrm{out}\right\rangle $ as tensor products of the
corresponding partial vacua in the five ranges $\Omega _{k}$, $k=1,...,5$.
The partial vacua are stable in $\Omega _{k}$, $k=1,2,4,5$, and the vacuum
instability with $\left\vert c_{v}\right\vert \neq 1$ is due to the partial
vacuum-to-vacuum transition amplitude formed in $\Omega _{3}$. In the range $%
\Omega _{3}$ operators of the number of final electrons and positrons are%
\begin{equation}
\hat{N}_{n}^{a}\left( \mathrm{out}\right) =\ ^{+}a_{n}^{\dagger }(\mathrm{out%
})\ ^{+}a_{n}(\mathrm{out}),\;N_{n}^{b}\left( \mathrm{out}\right) =\
_{+}b_{n}^{\dagger }(\mathrm{out})\ _{+}b_{n}(\mathrm{out}).  \label{Nop}
\end{equation}%
Using the linear canonical transformation between \textrm{in} and \textrm{out%
}-operators of creation and annihilation (see Eq. (7.4) in Ref. \cite%
{GavGi16}) one sees that the differential mean numbers of electrons and
positrons created from vacuum are presented by Eq. (\ref{7.5}). The
vacuum-to-vacuum transition amplitude, given by Eq. (7.21) in Ref. \cite%
{GavGi16}), is%
\begin{equation}
c_{v}=\langle 0,\mathrm{out}|0,\mathrm{in}\rangle =\prod\limits_{n\in \Omega
_{3}}g\left( ^{-}\left\vert _{-}\right. \right) g\left( ^{-}\left\vert
_{+}\right. \right) ^{-1}\,.  \label{cq30}
\end{equation}%
Then the probability for a vacuum to remain a vacuum can be presented by Eq.
(\ref{3.5}).

\section{Locally constant field approximation\label{Ap2}}

A new kind of a locally constant field approximation (LCFA) was formulated
in Ref. \emph{\cite{GavGitShi19b}. }Here we pretend to show that the density
of created pairs (\ref{3.41}) and the probability of the vacuum to remain a
vacuum (\ref{3.42}), obtained from exact equations for the slowly varying
field in the leading-term approximation, are in agreement with results
following in the framework of LCFA.

We call the electric field $E(x)$ a weakly inhomogeneous electric field on a
spatial interval $\Delta l$ if the following condition holds true:%
\begin{equation}
\left\vert \frac{\overline{\partial _{x}E(x)}\Delta l}{\overline{E(x)}}%
\right\vert \ll 1,\ \ \Delta l/\Delta l_{\mathrm{st}}^{\mathrm{m}}\gg 1\ ,
\label{sc.1}
\end{equation}%
where $\overline{E(x)}$ and $\overline{\partial _{x}E(x)}$ are the mean
values of $E(x)$ and $\partial _{x}E(x)$ on the spatial interval $\Delta l$,
respectively, and $\Delta l$ is significantly larger than the length scale $%
\Delta l_{\mathrm{st}}^{\mathrm{m}}$, which is%
\begin{equation}
\Delta l_{\mathrm{st}}^{\mathrm{m}}=\Delta l_{\mathrm{st}}\max \left\{
1,m^{2}/e\overline{E(x)}\right\} ,\ \ \Delta l_{\text{\textrm{st}}}=\left[ e%
\overline{E(x)}\right] ^{-1/2}\ .  \label{sc.2}
\end{equation}

Note that the length scale $\Delta l_{\mathrm{st}}^{\mathrm{m}}$ appears in
Eq.~(\ref{sc.1}) as the length scale when the perturbation theory with
respect to the electric field breaks down and the Schwinger
(nonperturbative) mechanism is primarily responsible for the pair creation.
In what follows, we show that this condition is sufficient. We are primarily
interested in strong electric fields, $m^{2}/e\overline{E(x)}\lesssim 1$. In
this case, the second inequality in Eq.~(\ref{sc.1}) is simplified to the
form $\Delta l/\Delta l_{\mathrm{st}}\gg 1$, in which the mass $m$ is
absent. In such cases, the potential of the corresponding electric step
hardly differs from the potential of a uniform electric field,%
\begin{equation}
U(x)=-eA_{0}(x)\approx U_{\mathrm{const}}(x)=e\overline{E(x)}x+U_{0},
\label{sc.3}
\end{equation}%
on the interval $\Delta l$, where $U_{0}$ is a given constant.

For an arbitrary weakly inhomogeneous strong electric field, in the
leading-term approximation, were derived universal formulas for the total
density of created pairs 
\begin{equation}
\rho ^{\mathrm{\Omega }}\approx \frac{J_{(d)}}{(2\pi )^{d-1}}\int_{x_{%
\mathrm{L}}}^{x_{\mathrm{R}}}dx\ eE(x)\int d\mathbf{p}_{\bot }N_{n}^{\mathrm{%
uni}},\ \ N_{n}^{\mathrm{uni}}=\exp \left[ -\pi \frac{\pi _{\bot }^{2}}{eE(x)%
}\right] \ ,  \label{np.7}
\end{equation}%
and an expression for the probability $P_{\mathrm{v}}$ given by Eq. (\ref%
{3.5}) of the vacuum to remain a vacuum,%
\begin{equation}
P_{\mathrm{v}}\approx \exp \left\{ -\frac{V_{\bot }TJ_{(d)}}{(2\pi )^{d-1}}%
\sum_{l=1}^{\infty }\int_{x_{\mathrm{L}}}^{x_{\mathrm{R}}}dx\frac{\left[
eE(x)\right] ^{d/2}}{l^{d/2}}\exp \left[ -\pi \frac{lm^{2}}{eE(x)}\right]
\right\} \ ;  \label{np.8}
\end{equation}%
see Ref. \cite{GavGitShi19b}. In Eqs. (\ref{np.7}) and (\ref{np.8})
integration limits are specified over the region $S_{\mathrm{int}}=(x_{%
\mathrm{L}},x_{\mathrm{R}})$, in which the electrical the field is not zero.
In our case $x_{\mathrm{L}}=-\infty $, $x_{\mathrm{R}}=+\infty $.

Let's compare Eqs. (\ref{np.7}) and (\ref{np.8}) with the results obtained
above in (\ref{3.41}). To do this, let's represent (\ref{np.7}) in the form%
\begin{eqnarray}
&&\rho ^{\mathrm{\Omega }}\approx \frac{J_{\left( d\right) }}{\left( 2\pi
\right) ^{d-1}}\int d\mathbf{p}_{\bot }\left( J_{p_{\bot }}^{+}+J_{p_{\bot
}}^{-}\right) \,,  \notag \\
&&J_{p_{\bot }}^{+}=\int_{x_{\mathrm{M}}}^{\infty }dx\left[ eE\left(
x\right) \right] N_{n}^{\mathrm{univ}}=-e\int_{x_{\mathrm{M}}}^{\infty
}dA_{0}(x)N_{n}^{\mathrm{univ}}\,\nonumber\\
&&J_{p_{\bot }}^{-}=\int_{-\infty }^{x_{%
\mathrm{M}}}dx\left[ eE\left( x\right) \right] N_{n}^{\mathrm{univ}%
}=-e\int_{x_{\mathrm{M}}}^{\infty }dA_{0}(x)N_{n}^{\mathrm{univ}}\ .
\label{np.10}
\end{eqnarray}

Note that the functions $A_{0}(x)$ and $E(x)$ can be related to each other
using the cubic equation%
\begin{equation}
y^{3}-y-\frac{2}{3\sqrt{3}(q+1)}=0,\ \ y=-A_{0}(x)/(E_{0}\sigma ),\ \
q=\left( 3\sqrt{3}\frac{E(x)}{E_{0}}\right) ^{-1}-1\ .  \label{np.11}
\end{equation}%
We can express $A_{0}\left( x\right) $ as a function of the field $E\left(
x\right) $ or as a function of the variable $q$ using solutions of equation (%
\ref{np.11}). This equation has three real solutions,%
\begin{align}
& y_{1}=\frac{2}{\sqrt{3}}\cos \frac{\alpha \left( q\right) }{3},\ y_{2}=-%
\frac{2}{\sqrt{3}}\cos \left[ \frac{\alpha \left( q\right) }{3}+\frac{\pi }{3%
}\right] \ ,  \notag \\
& y_{3}=-\frac{2}{\sqrt{3}}\cos \left[ \frac{\alpha \left( q\right) }{3}-%
\frac{\pi }{3}\right] ,\ \alpha \left( q\right) =\arccos \left[ \left(
q+1\right) ^{-1}\right] \ ,  \label{gav4}
\end{align}%
see, e.g., \cite{Korn}. Since $A_{0}\left( x\right) $ is negative, only the
solutions $y_{2,3}$ are relevant. One can see that for solutions $y_{2,3}$
the differential $dA_{0}\left( x\right) $ takes the form:%
\begin{eqnarray}
&&dA\left( x\right) =-2(E_{\max }\sigma )\sin \left[ \frac{\alpha \left(
q\right) }{3}+\frac{\pi }{3}\right] \frac{\left( q+1\right) ^{-2}}{\sqrt{%
1-\left( q+1\right) ^{-2}}}dq,\ \ x>0\ ,  \notag \\
&&dA\left( x\right) =-2(E_{\max }\sigma )\sin \left[ \frac{\alpha \left(
q\right) }{3}-\frac{\pi }{3}\right] \frac{\left( q+1\right) ^{-2}}{\sqrt{%
1-\left( q+1\right) ^{-2}}}dq\ \ x\leq 0\ ,  \notag \\
&&E(x)=-dA(x)=\frac{E_{\max }}{q+1}\ .  \label{np.12}
\end{eqnarray}%
Passing from the integration over $x$ to the integration over the parameter $%
q$ in Eq. (\ref{np.10}), we find:%
\begin{equation}
J_{p_{\bot }}^{\pm }=I_{p_{\bot }}^{\pm }\ ,  \label{gav6}
\end{equation}%
where the quantities $I_{p_{\bot }}^{\pm }$ are given by Eq. (\ref{3.39}).

It follows from Eq. (\ref{gav6}) that the density of the created pairs (\ref%
{np.7}) and the probability of the vacuum to remain a vacuum (\ref{np.8})
obtained with the help of the approximation for weakly inhomogeneous strong
electric field coincide with expressions (\ref{3.41}) and (\ref{3.42}),
respectively.

\section{Some properties of hypergeometric functions\label{A1}}

The hypergeometric function $F\left( a,b,c;z\right) =\,_{2}F_{1}\left(
a,b,c;z\right) $ (here and in what follows it is supposed that parameters $a$
and $b$ are not equal to $0,-1,-2$, $\ldots $) is defined by series%
\begin{equation}
F\left( a,b,c;z\right) =\sum_{n=0}^{+\infty }\frac{\left( a\right)
_{n}\left( b\right) _{n}}{\left( c\right) _{n}}\frac{z^{n}}{n!}=\frac{\Gamma
\left( c\right) }{\Gamma \left( a\right) \Gamma \left( b\right) }%
\sum_{n=0}^{+\infty }\frac{\Gamma \left( a+n\right) \Gamma \left( b+n\right) 
}{\Gamma \left( c+n\right) }\frac{z^{n}}{n!},\ \ \left\vert z\right\vert <1.
\label{A01}
\end{equation}%
Note that in the solutions (\ref{2.13}) and (\ref{2.14}) the arguments $%
1-\xi ^{-1}$ and $\xi ^{-1}$ in the corresponding hypergeometric functions
are less than unity and the series (\ref{A01}) converges.

At $\left\vert z\right\vert =1$ the series (\ref{A01}) converges absolutely
when $\mathrm{Re}\left( c-a-b\right) >0$. The integral representation%
\begin{equation}
F\left( a,b,c;z\right) =\frac{\Gamma \left( c\right) }{\Gamma \left(
b\right) \Gamma \left( c-b\right) }\int_{0}^{1}t^{b-1}\left( 1-t\right)
^{c-b-1}\left( 1-zt\right) ^{-a}dt,\ \ \left( \mathrm{Re }c>\mathrm{Re }
b>0\right)  \label{A02}
\end{equation}%
gives an analytical continuation for the function $F\left( a,b,c;z\right) $
to the complex $z$-plane with a cut along the real axis from $1$ to $\infty $
(since the right-hand side is an unambiguous analytic function in the domain 
$\left\vert \arg \left( 1-z\right) \right\vert \leq \pi $). From the
integral representation (\ref{A02}) it is easy to see that $%
\lim_{z\rightarrow 0}F\left( a,b,c;z\right) =1.$ The formula for
differentiating the hypergeometric function has the form: 
\begin{equation}
\frac{d}{dz}F\left( a,b,c;z\right) =\frac{ab}{c}F\left( a+1,b+1,c+1;z\right)
.  \label{AdF}
\end{equation}

It is follows from (\ref{A02}) that 
\begin{equation}
F\left( a,b,c;z\right) =\left( 1-z\right) ^{c-a-b}F\left( c-a,c-b,c;z\right)
,\ \ \left\vert z\right\vert <1.  \label{A04}
\end{equation}

Hypergeometric function can be transformed as 
\begin{eqnarray}
 F\left( a,b,c;z\right) &=&\frac{\Gamma \left( c\right) \Gamma \left(
c-a-b\right) }{\Gamma \left( c-a\right) \Gamma \left( c-b\right) }F\left(
a,b,a+b-c+1;1-z\right)  \notag \\
&+&\left( 1-z\right) ^{c-a-b}\frac{\Gamma \left( c\right) \Gamma \left(
a+b-c\right) }{\Gamma \left( a\right) \Gamma \left( b\right) }F\left(
c-a,c-b,c-a-b+1;1-z\right)\,,\nonumber\\
&&\left( \left\vert \arg \left( 1-z\right)
\right\vert <\pi \right) ,  \label{A05a}
\end{eqnarray}

The hypergeometric equation in its general form, 
\begin{equation}
z\left( 1-z\right) w^{\prime \prime }\left( z\right) +\left[ c-\left(
a+b+1\right) z\right] w^{\prime }\left( z\right) -abw\left( z\right) =0,
\label{A06}
\end{equation}%
has three regular singular points $z=0,1,\infty $. When none of the numbers $%
c$, $c-a-b$, $a-b$ is integer, the general solution $w\left( z\right) $ of
the hypergeometric equation (\ref{A06})\ can be obtained as 
\begin{eqnarray}
&&w\left( z\right) =c_{1}w_{1}\left( z\right) +c_{2}w_{2}\left( z\right) ,\ 
\text{ }z\rightarrow 0,  \notag \\
&&w\left( z\right) =c_{1}w_{3}\left( z\right) +c_{2}w_{4}\left( z\right) ,\
\ z\rightarrow 1,  \notag \\
&&w\left( z\right) =c_{1}w_{5}\left( z\right) +c_{2}w_{6}\left( z\right) ,\
\ z\rightarrow \infty \text{.}  \label{A07set}
\end{eqnarray}%
where $c_{1}$ and $c_{2}$ are some constants, and the functions $w_{j}\left(
z\right) $, $j=1,\ldots ,6$, have the form: 
\begin{eqnarray}
&&w_{1}\left( z\right) =F\left( a,b,c;z\right) ,\ \ w_{2}\left( z\right)
=z^{1-c}F\left( a-c+1,b-c+1,2-c;z\right) ,  \notag \\
&&w_{3}\left( z\right) =F\left( a,b,a+b+1-c;1-z\right) ,\   \notag \\
&&w_{4}\left( z\right) =\left( 1-z\right) ^{c-a-b}F\left(
c-b,c-a,c-a-b+1,1-z\right) ,  \notag \\
&&w_{5}\left( z\right) =z^{-a}F\left( a,a-c+1,a-b+1,z^{-1}\right) ,\   \notag
\\
&&w_{6}\left( z\right) =z^{-b}F\left( b,b-c+1,b-a+1,z^{-1}\right) .
\label{A08}
\end{eqnarray}%
The Kummer relations and for the hypergeometric equation \cite{Bateman}
allow us to represent the functions $w_{1}\left( z\right) $ and $w_{2}\left(
z\right) $ via the functions $w_{3}\left( z\right) $ and $w_{4}\left(
z\right) $,%
\begin{eqnarray}
w_{1}\left( z\right) &=&e^{i\pi \left( 2\alpha _{1}-b\right) }\frac{\Gamma
\left( 2\left( \alpha _{1}+1\right) -a-b\right) \Gamma \left( b-a+1\right) }{%
\Gamma \left( 2-a\right) \Gamma \left( 2\alpha _{1}-a+1\right) }w_{4}\left(
z\right)  \notag \\
&-&e^{i\pi \left( 2\alpha _{1}-a\right) }\frac{\Gamma \left( 2\left( \alpha
_{1}+1\right) -a-b\right) \Gamma \left( a-b-1\right) }{\Gamma \left(
1-b\right) \Gamma \left( 2\alpha _{1}-b\right) }w_{3}\left( z\right) , 
\notag \\
w_{2}\left( z\right) &=&e^{i\pi \left( a-1\right) }\frac{\Gamma \left(
a+b-2\alpha _{1}\right) \Gamma \left( b-a+1\right) }{\Gamma \left( b-2\alpha
_{1}+1\right) \Gamma \left( b\right) }w_{4}\left( z\right)  \notag \\
&+&e^{i\pi b}\frac{\Gamma \left( a+b-2\alpha _{1}\right) \Gamma \left(
a-b-1\right) }{\Gamma \left( a-2\alpha _{1}\right) \Gamma \left( a-1\right) }%
w_{3}\left( z\right) .  \label{cm1}
\end{eqnarray}

\end{document}